\DocumentMetadata{}
\documentclass[manuscript,screen,prologue,table,xcdraw]{acmart}
\usepackage{caption}
\usepackage{subcaption}
\usepackage{array}
\usepackage{multirow}
\usepackage{rotating}
\usepackage{float}
\usepackage{MnSymbol}

\usepackage{booktabs}
\usepackage{hyperref}

\usepackage{soul} % \hl{} highlighting

\newif\ifshowchanges
\showchangestrue
\showchangesfalse % ML 1/17/24 after submission of revision

\ifshowchanges
    \usepackage{changes} 
\else % \ifshowchanges
    \newcommand{\added}[1]{#1}
    \newcommand{\deleted}[1]{}

\fi
\AtBeginDocument{%
  \providecommand\BibTeX{{%
    \normalfont B\kern-0.5em{\scshape i\kern-0.25em b}\kern-0.8em\TeX}}}
    
\usepackage{booktabs}
\usepackage{graphicx}
\usepackage{longtable}
\usepackage{multirow}

\setcopyright{rightsretained}
\copyrightyear{2018}
\acmYear{2018}
\acmDOI{XXXXXXX.XXXXXXX}

\acmConference[Conference acronym 'XX]{Make sure to enter the correct
  conference title from your rights confirmation emai}{June 03--05,
  2018}{Woodstock, NY}
\begin{document}
%TC:ignore
%%
%% The "title" command has an optional parameter,
%% allowing the author to define a "short title" to be used in page headers.
\title[Co-Designing with Fact-checkers using Matchmaking for AI]{Human-centered NLP Fact-checking: Co-Designing with Fact-checkers using Matchmaking for AI} 
%\subtitle{}

%%
%% The "author" command and its associated commands are used to define
%% the authors and their affiliations.
%% Of note is the shared affiliation of the first two authors, and the
%% "authornote" and "authornotemark" commands
%% used to denote shared contribution to the research.
\author{Houjiang Liu}
\authornote{Three authors contributed equally to this research.}
% \email{trovato@corporation.com}
% \orcid{1234-5678-9012}
% \email{webmaster@marysville-ohio.com}
\email{liu.ho@utexas.edu}
\orcid{0000-0003-0983-6202}
\affiliation{%
  \institution{School of Information, University of Texas at Austin}
  % \department{School of Information}
  % \streetaddress{P.O. Box 1212}
  \city{Austin}
  \state{Texas}
  \country{USA}
  % \postcode{43017-6221}
}

\author{Anubrata Das}
\email{anubrata.das@utexas.edu}
\orcid{0000-0002-5412-6149}
\authornotemark[1]
\affiliation{%
  % \department{}
  \institution{School of Information, University of Texas at Austin}
  % \department{School of Information}
  % \streetaddress{P.O. Box 1212}
  \city{Austin}
  \state{Texas}
  \country{USA}
  % \postcode{43017-6221}
}

\author{Alexander Boltz}
\email{aboltz@uw.edu}
\orcid{0000-0003-2666-1838}
\authornotemark[1]
\affiliation{%
  \institution{Human Centered Design \& Engineering, University of Washington}
  % \streetaddress{P.O. Box 1212}
  \city{Seattle}
  \state{Washington}
  \country{USA}
  % \postcode{43017-6221}
}

\author{Didi Zhou}
\email{didizhou@utexas.edu}
\orcid{0000-0002-9637-6154}
\affiliation{%
  \institution{School of Information, University of Texas at Austin}
  % \department{School of Information}
  % \streetaddress{P.O. Box 1212}
  \city{Austin}
  \state{Texas}
  \country{USA}
  % \postcode{43017-6221}
}

\author{Daisy Pinaroc}
\email{drp2767@utexas.edu}
\orcid{0009-0002-0448-8851}
\affiliation{%
  \institution{School of Information, University of Texas at Austin}
  % \department{School of Information}
  % \streetaddress{P.O. Box 1212}
  \city{Austin}
  \state{Texas}
  \country{USA}
  % \postcode{43017-6221}
}

\author{Matthew Lease}
\email{ml@utexas.edu}
\orcid{0000-0002-0056-2834}
\affiliation{%
  \institution{School of Information, University of Texas at Austin}
  % \department{School of Information}
  % \streetaddress{P.O. Box 1212}
  \city{Austin}
  \state{Texas}
  \country{USA}
  % \postcode{43017-6221}
}

\author{Min Kyung Lee}
\email{minkyung.lee@austin.utexas.edu}
\orcid{0000-0002-2696-6546}
\affiliation{%
  \institution{School of Information, University of Texas at Austin}
  % \department{School of Information}
  % \streetaddress{P.O. Box 1212}
  \city{Austin}
  \state{Texas}
  \country{USA}
  % \postcode{43017-6221}
}

%%
%% By default, the full list of authors will be used in the page
%% headers. Often, this list is too long, and will overlap
%% other information printed in the page headers. This command allows
%% the author to define a more concise list
%% of author names for this purpose.
\renewcommand{\shortauthors}{Houjiang Liu et al.}

%%
%% The abstract is a short summary of the work to be presented in the
%% article.
\begin{abstract}
% ML 1/14/24: cutting first sentence as suggested earlier
%\replaced{One of the main challenges in human fact-checking is its limited efficiency and scalability, especially given the overwhelming amount of false information.}{A key challenge in professional fact-checking is its limited scalability in relation to the magnitude of false information.}
% in relation to the magnitude of false information. 
While many Natural Language Processing (NLP) techniques have been proposed for fact-checking, both academic research and fact-checking organizations report limited adoption of such NLP work due to poor alignment with fact-checker practices, values, and needs. To address this, we investigate a co-design method, \textit{Matchmaking for AI}, to enable fact-checkers, designers, and NLP researchers to collaboratively identify what fact-checker needs should be addressed by technology, and to brainstorm ideas for potential solutions. Co-design sessions we conducted with 22 professional fact-checkers yielded a set of 11 design ideas that offer a ``north star'', integrating fact-checker criteria into novel NLP design concepts. These concepts range from pre-bunking misinformation, efficient and personalized monitoring misinformation, proactively reducing fact-checker potential biases, and collaborative writing fact-check reports. Our work provides new insights into both human-centered fact-checking research and practice and AI co-design research.

\end{abstract}
%TC:endignore

%%
%% The code below is generated by the tool at http://dl.acm.org/ccs.cfm.
%% Please copy and paste the code instead of the example below.
%%


%%
%% Keywords. The author(s) should pick words that accurately describe
%% the work being presented. Separate the keywords with commas.

%%
%% This command processes the author and affiliation and title
%% information and builds the first part of the formatted document.
\maketitle

\section{Introduction} \label{sec:intro}

The massive scale of intentional {\em disinformation} and inadvertent {\em misinformation} threaten modern society in many areas, such as law and order \citep{Wardle2019-lv}, politics \citep{Weeks2015-mq}, and healthcare \citep{Suarez-Lledo2021-aj}. To combat this, journalists and fact-checkers perform a key societal function by debunking fake news, online rumors, and conspiracy \citep{Dobbs2012-su, Jack2017-cn}.  Today there are over 400 teams of journalists and researchers in over 105 countries with 378 known, active fact-checking projects \cite{Stencel2022-ab}. While such human fact-checking has proven to be effective in terms of various measures \citep{Porter2021-nk}, fact-checking remains a largely manual affair today, limiting the scale of its effective reach and practical impact \citep{micallef2022true, Procter2023-rr}. 

To help address this scalability challenge, many Natural Language Processing (NLP) techniques and tools have been proposed for fact-checking, across tasks such as claim detection, evidence retrieval, and claim verification  \citep{Adair2020, Funke2017-yj, ThePoynterInstitute2021}.  Beyond NLP, additional AI techniques have been proposed for fact-checking multi-modal content (images, videos, and audio) \citep{Alam2021-zd, Cao2020-oi,Maros2021-le}. However, despite the clear need for greater scalability and this ever-growing body of AI work on fact-checking, to date there has been only limited adoption of AI-based methods and tools for fact-checking. As highlighted by both academic research \citep{micallef2022true, graves2018understanding, juneja2022human} and fact-checking organization reports \citep{ArnoldPh:online}, this stems from insufficient alignment with fact-checker needs, practices, and values. In part, simplified problem formulations that permit full automation neglect important aspects of real-world fact-checking, which remains a highly complex process requiring subjective judgment and human knowledge to corroborate claims based on local contexts \citep{graves2018understanding, micallef2022true}. Relatedly, AI-based fact-checking driven by technological challenges of cutting-edge AI neglects the need for design studies to better surface real fact-checker needs and translate these into human-centered AI tools responsive to these needs \citep{Das2023-tq, nakov2021automated}.

In this work, we use co-design to engage fact-checkers directly in designing human-centered NLP tools to meet their needs. To limit the scope, we restrict our study to NLP technologies related to language-based fact-checking, omitting broader AI technologies related to multi-modal fact-checking. Fact-checkers, designers, and NLP researchers translate their needs through co-design into design ideas informed by state-of-the-art NLP techniques for fact-checking. 

In particular, we propose \textbf{\textit{Matchmaking for AI}} to conduct this co-design, extending \citet{Bly1999-ar}'s earlier {\em Matchmaking} concept to an AI co-design process. {\em Matchmaking for AI} distills AI techniques into AI probes---abstracted visual concepts and forms---that are easy for non-AI experts to understand and provide a step-by-step ideation process to map state-of-the-art AI techniques into user activities wherein those techniques may have the most impact. Prior work in human-centered AI has mainly incorporated stakeholder feedback in two ways: 1) enhancing existing design with human values \citep{Zhu2018-di, Lee2019-kg, Zhou2020-ww, Subramonyam2021-yn}, and 2) brainstorming with stakeholders to translate their values into high-level design guidelines \citep{zhang2022algorithmic, harrington2021eliciting, Komatsu2020-gl, skinner2020children}. However, there appears to be a potential gap -- while numerous frameworks have been developed to integrate human values into AI development, brainstorming the ``right'' design concepts with stakeholders that are novel and feasible remains a challenge. Additionally, existing AI co-design practices for brainstorming ideas tend to be speculative and may not consistently focus on the practical feasibility of these ideas. Thus, {\em Matchmaking for AI} empowers stakeholders to co-design a suite of workable ideas that support their workflow. 

Using {\em Matchmaking for AI}, we conducted co-design sessions with 22 professional fact-checkers, yielding 11 novel design ideas (to the best of our knowledge). Compared to existing NLP fact-checking that focused on claim selection and verification, our ideas point to a broader set of challenges across the fact-checking workflow and fact-checker goals. The ideas assist in information searching, processing, and writing tasks for efficient and personalized fact-checking; help fact-checkers proactively prepare for future misinformation; monitor their potential biases; and support internal organization collaboration. In addition, by using AI probes to teach non-technical stakeholders ``just enough'' AI to collaborate effectively with specialists, {\em Matchmaking for AI} helped foster effective communications with AI researchers, to provide concrete design suggestions and relevant technical details, so that participants could more readily brainstorm feasible AI design ideas. 

Our work contributes to the literature on human-centered fact-checking in Human-computer Interaction (HCI) and NLP by offering a suite of fact-checker-designed tool ideas that can guide future NLP-based fact-checking research. The ideas could lead to technology development with greater potential for adoption by professional fact-checkers in practice. Our work also contributes to the literature on AI co-design by offering the use case of {\em Matchmaking for AI} with fact-checkers, which facilitates co-design by fact-checkers, designers, and NLP researchers.

The remainder of this article is organized as follows. We first review key related work in Section \ref{sec:related work}. Then in Section \ref{sec:studydesign}, we describe Matchmaking for AI in greater detail, study protocol, recruitment of fact-checkers, and methods used to analyze co-design outcomes. In Section \ref{sec:results}, we present 11 ideas co-designed by fact-checkers. Next, in Section \ref{sec:discussion}, we reflect on its benefits, efficacy, limitations, and future improvements. Finally, we conclude with Section \ref{sec:conclusion}.

\section{Related Work} \label{sec:related work}

\subsection{Professional Fact-checking Today and Opportunities for Mixed-initiative Workflows} \label{subsec:human_fact-checking}

Fact-checking practices exhibit some common patterns and workflows across different organizations. For example, \citet{graves2017anatomy} synthesizes a standard fact-checking process into five stages: 1) choosing claims to check, 2) contacting the speaker, 3) tracing false claims, 4) working with experts, and 5) writing the fact-check. Building on \citet{graves2017anatomy}'s work, \citet{micallef2022true} grounds a more detailed workflow by interviewing fact-checkers worldwide. \citet{juneja2022human} explore both the human and technical infrastructure of fact-checking among different stakeholder groups (e.g., editors, fact-checkers, investigators and researchers, social media managers, and advocates). As reported by these studies, a key challenge is the difficulty for human fact-checkers in combating the massive scale of online misinformation.

In other work, a growing body of NLP research has sought to automate different aspects of fact-checking \citep{guo2022survey}. To date, most attention has been directed toward: 1) detecting claims and prioritizing claim checkworthiness \citep{Shaar2021-jk, Hassan2017TowardAF}, 2) detecting previously fact-checked claims \citep{Kazemi2021-vg}, and 3) assessing claims given textual evidence \citep{shahi2021overview, Sarrouti2021-cl}. However, as reported by recent academic studies \cite{micallef2022true, Procter2023-rr} and organization reports \cite{ArnoldPh:online}, most existing computational tools are still ``fragmented'' \cite{micallef2022true} and have ``limited scope and use of custom solutions'' \cite{micallef2022true, beers2020examining}.

% ML 1/14/24 cut
%Below, we briefly review off-the-shelf computational tools that are \replaced{powered}{empowered} by these NLP techniques.

Two popular claim detection tools include \textit{ClaimBuster}\citep{Hassan2017-pt}, an open-sourced API to extract claims from paragraphs, and \textit{Full Fact Alpha}\footnote{\url{https://fullfact.org/about/ai/}},
which enables fact-checkers to search and filter checkable claims from news articles. Claim detection is also incorporated into speech recognition tools \citep{Adair2020}, social media monitoring tools\footnote{\url{https://www.facebook.com/formedia/mjp/programs/third-party-fact-checking}}, and tip-line services\footnote{\url{https://meedan.com/post/covid-19-whatsapp-bot-for-fact-checkers}}. However, these tools struggle to correctly prioritize claims in alignment with human fact-checker's definitions of ``check-worthiness'' \citep{ArnoldPh:online, Sehat2023-xa}. To investigate claims, fact-checkers use off-the-shelf search engines, such as Google and Bing, to surface related articles, but most topically relevant pages retrieved from these search engines usually do not contain useful content for fact-checking \citep{beers2020examining, ArnoldPh:online}. For claim verification, existing tools include (but are not limited to): 1) verifying COVID-19 related claims based on official datasets \citep{dong2020interactive}; 2) checking statistical macroeconomic claims by referring to government reports and national statistics \citep{ThePoynterInstitute2021}; and 3) checking scientific claims based on academic paper abstracts \cite{wadden2020fact}. However, due to their limited scope, fact-checkers have not widely adopted these auto-verification tools. 

% ML 1/15/23 these sentences are kinda weird; commenting out
%Although computational tools can improve fact-checker productivity to some degree, the goal of AI automation is not to simply emulate human work. Even if certain domain-specific AI techniques are labeled as likely to do so (e.g., claim detection and verification), these models are primarily used to filter large irrelevant information instead of replacing fact-checker work. 
As reported by empirical studies \citep{graves2018understanding, micallef2022true, juneja2022human, Procter2023-rr}, fact-checking remains a complex practice that inherently requires subjective judgment and human knowledge \citep{ArnoldPh:online, graves2017anatomy, micallef2022true}, encompassing a variety of tasks that remain beyond the capabilities of even state-of-the-art AI. For this reason, fact-checking tools appear most likely to impact practice if researchers and designers seek to develop assistive technologies that augment, enhance, and accelerate the work of professional fact-checkers, rather than seeking to completely automate fact-checking. We thus envision a \textit{mixed-initiative process} \citep{Horvitz1999-wp}: fact-checking tasks are strategically divided into more fine-grained sub-tasks operated by human fact-checkers and AI \citep{Lease2020-xa, Nguyen2018believeornot, nakov2021automated}. Human-centered design and co-design are considered vital methodologies for combining the strengths of both human effort and AI to create complementary hybrid solutions \citep{Delgado2021-gk, Das2023-tq}.

\subsection{Human-Centered Design and Co-Design for AI} \label{subsec:co-designAI}

Engaging stakeholders during AI development and design has increasingly piqued scholars’ attention among HCI and AI communities \citep{Zytko2022-ge, Delgado2021-gk}. Empirical studies from different AI application domains, such as AI-assisted public and clinical decision-making \citep{Yang2019-kw, Sun2019-vj, Starke2022-hc}, online advertising \citep{Ali2019-hl}, as well as automated fact-checking \citep{micallef2022true, juneja2022human, Procter2023-rr}, reported that the lack of stakeholder engagement in AI development might produce fairness-related harm for stakeholders at risk of poor AI performance. Additionally, the implementation and design of AI tools might fail to meet real stakeholder needs because they do not adapt to the socio-technical context and norms within specialized domains \citep{Komatsu2020-gl}.

To address this, various human-centered approaches --- including frameworks, methods, and design materials --- have been proposed to broaden stakeholder participation in AI design and development. For example, \citet{Zhu2018-di} introduced Value-Sensitive Algorithm Design to create ``intelligent socialization algorithms for WikiProjects in Wikipedia.'' \citet{Lee2019-kg} developed the WeBuildAI framework to design algorithms for ``on-demand food donation transportation services'' that balance stakeholder values. Also,  HCI scholars propose using \textit{AI as a design material} \citep{dove2017ux}, in order to integrate AI elements into a traditional UX design process. A variety of design materials have been created to pair participatory AI with non-expert AI stakeholders, including designers \citep{Zhou2020-ww, Subramonyam2021-yn} and end-users \citep{Kuo2023-ms}. For example, \citet{Zhou2020-ww} designed a user journey canvas that was incorporated with an AI development process (data annotation, model training, and inference). Designers can use this canvas to specify different elements for an AI design idea. Like \citet{Zhou2020-ww}, \citet{Subramonyam2021-yn} created data-persona cards, model API cards, and explainability design guidelines, to engage AI practitioners (designers and engineers) to co-create AI experiences. \citet{Kuo2023-ms} also mapped an AI development lifecycle into comic boards, which help front-line workers understand how AI is developed and share detailed suggestions for designing AI systems. 

Note that the aforementioned human-centered design methods are commonly employed to enhance design details by incorporating human values, particularly when a design concept is already established. In contrast, co-design is used to assist stakeholders in brainstorming ideas in situations where the AI design landscape is under-explored, translating social and ethical values as design guidance or distilling high-level user needs. For example, it has been used to co-imagine ``tech futures'' for a Utopian city with Black youth \citep{harrington2021eliciting}; to understand how children imagine a fair AI librarian would behave \citep{skinner2020children}; to uncover values to inform AI design for journalism \citep{Komatsu2020-gl}; to imagine new rideshare platforms that consider driver well-being \citep{zhang2022algorithmic}; and to raise awareness of algorithmic controls on social media feeds \citep{Storms2022-nq}.

\citet{sanders2008co} define co-design as a co-creation activity, involving expert users who actively contribute to brainstorming new design ideas and informing future product development. This suggests that these users possess some level of technical expertise. As they begin to use existing tools with an innovative and expert mindset, co-design can effectively integrate their insights into the creation of more useful tools \cite{sanders2008co, Visser2005-nt}. However, by reviewing recent AI co-design practices, we learn that as most stakeholders are not familiar with AI, AI co-design often becomes speculative, primarily serving as a method for human value solicitation, rather than yielding specific design decisions with concrete AI technology being used \citep{Delgado2021-gk}. Thus, there appears a potential gap in the existing AI design paradigm -- while numerous human-centered frameworks have been created for incorporating human values into AI development, identifying the ``right'' design concepts upfront, particularly those that are innovative yet feasible for stakeholder-centered AI design remains a challenge. Additionally, existing AI co-design practices for brainstorming ideas tend to be speculative and may not consistently focus on the practical feasibility of these ideas.

% ML 1/15/24: the transition from related work to this deep discussion of ours seems abrupt, so adding this subsection to make that transition more clear
\subsection{Design through Matchmaking} \label{subsec:mmAI}

In this work, we seek to bridge the above gap by empowering stakeholders to brainstorm AI ideas that are both novel and feasible --- fact-checkers and AI researchers have deeper-level conversations to explore appropriate design choices with a better adoption of existing AI techniques, so that these design ideas are practically useful for fact-checkers and more likely to be developed in the real world. In particular, we introduce \textbf{\textit{Matchmaking for AI}}, extending \citet{Bly1999-ar}'s Matchmaking concept as an AI co-design process, which maps AI techniques to user activities wherein those techniques may have the most impact. \citet{Kensing2012-ni} point out that to foster effective co-design, mutual learning is a necessary step for different stakeholders by sharing and synthesizing their domain knowledge to brainstorm novel design solutions. Different from studies that use co-design in a speculative approach without considering technological capabilities upfront, Bly and Churchill implement matchmaking with a series of mutual learning steps (described in Section \ref{subsec:matchmaking}), which help non-technical experts identify ``technology affordance'' and map this affordance to their work activities that can be easily supported, thus producing feasible design choices.

Inspired by this original process and an adapted version developed by \citet{Van_Dijk2019-pm}, our extended \textit{Matchmaking for AI} aims to foster such learning experiences between stakeholders and AI researchers. For example, by better understanding AI techniques, stakeholders could be more knowledgeable to inform what and how AI could augment domain-specific tasks instead of emulating human work \citep{Shneiderman2020-yb}. Meanwhile, AI researchers could become more creative to innovate domain-specific AI techniques that situate stakeholder needs \citep{Das2023-tq, nakov2021automated}.

\section{Study Design} \label{sec:studydesign}

As prior human-centered design helps produce sufficient design details and traditional co-design facilitates idea brainstorming, \textit{Matchmaking for AI} leverages synergistic interactions between stakeholder expertise and their understanding of AI capabilities to brainstorm feasible AI ideas. Our process actively familiarizes fact-checkers with fine-grained NLP techniques (e.g., classification, clustering, and summarization). With this learning experience, fact-checkers can envision how NLP might be usefully and realistically applied in their workflow and brainstorm how they can collaboratively work with AI automation across different fact-checking tasks.

In Section \ref{subsec:matchmaking}, we first describe the idea of Matchmaking for AI, including its original concept and our adaptation. Next, Section \ref{subsec:protocol} presents the study protocol and materials used in our workshops with fact-checkers. We then describe the recruitment in Section \ref{subsec:recruitment}, followed by the qualitative coding method used to analyze the workshops in Section \ref{subsec:analysis}.

\subsection{Matchmaking for AI} \label{subsec:matchmaking}

\citet{Bly1999-ar} propose ``Matchmaking'' as a co-design concept to ``incorporate user domain knowledge into early design when a technology prototype already exists''. The original concept includes four steps: 1) describing technology capabilities; 2) mapping those capabilities to associated work activities; 3) identifying work domains and specific tasks; and 4) verifying whether these tasks match technology capabilities. For the 2nd step, \citet{Van_Dijk2019-pm} argue that the mapping results of technology capabilities vs. work activities are too vast to be usefully navigated. To address this, they suggest that beyond merely mapping capabilities to activities, designers should further assess the rate at which mappings would be most useful. While prior matchmaking activities helped to identify stakeholders who might benefit from the new technology, designers, and developers still took an active role. We argue that this does not exploit the advantage of co-design, which democratizes the existing power structures of product development by empowering end-users with more autonomy to express their design choices. Thus, our \textit{Matchmaking for AI} inherits the original mapping activity for discovering design spaces where existing techniques can have the most impact but empowers stakeholders to do so by eliciting their expertise, passion, and creativity. 

\textit{Matchmaking for AI} seeks to create ``workable ideas'' that are both grounded in participant needs and values, as well as informed by their understanding of AI capabilities. This is expected to produce results that translate high-level themes (e.g., user needs or design guidelines) into clear design concepts that are more feasible for implementation. However, this does not mean that one could simply use off-the-shelf AI models for practical implementation. Instead, the goal is to identify ``right'' design concepts that would serve as a ``north star'' for designers and NLP researchers.
% to apply the human-centered framework they proposed to bring these ideas into practice.
% By reviewing previous AI studies that incorporate stakeholder feedback into the AI development (Section \ref{subsec:co-designAI}), we realized that ``workable ideas'' they have set up at the beginning do not necessarily represent ideas that can be directly implemented with off-the-shelf models. 
To generate such outcomes, we tailor a three-step matchmaking process that includes the following five aspects (a-e): a) need assessment, b) specific gaps identified from the existing tools and related prior work in the literature, c) tool features desired by our participants, d) mapping to existing AI techniques, and e) potential technical challenges for implementation. We discuss each of these three steps below.

\subsubsection{Mapping Stakeholder Domain Expertise and AI Needs (Step 1)} \label{subsec:domain_expertise}

\begin{figure}[!t]
    \centering
    \includegraphics[width=0.85\textwidth]{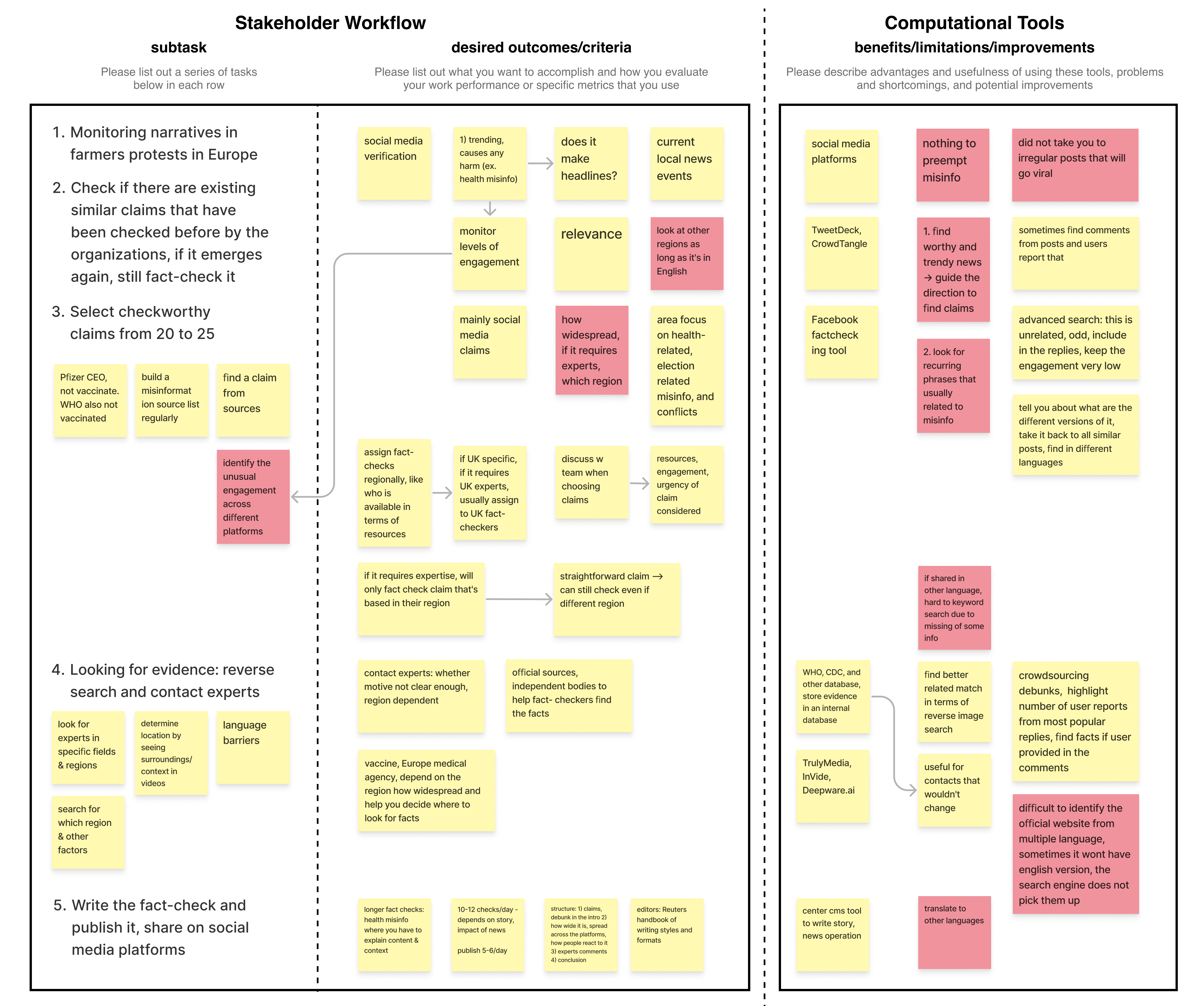}
    \caption{The \textit{Domain Specialty Canvas}: fact-checkers and design facilitators map out fact-checking workflow and computational tools. While fact-checkers were thinking aloud, designers facilitated distilling important concepts and recorded them (i.e., yellow sticky notes) in the canvas, followed by asking fact-checkers to affirm, revise, or add new content (i.e., red sticky notes).}
    \label{fig:stakeholderspecialties}
\end{figure}

A growing body of empirical work in human-centered AI has called for more extensive research on aligning AI capabilities with domain expertise \citep{Kohli2018-bx, Yang2019-kw, Cai2019-xb, Beede2020-uf}. Such alignments help designers and AI practitioners explore appropriate design spaces before developing AI applications. Our \textit{Matchmaking for AI} starts with capturing details of stakeholder domain expertise and their AI needs. Inspired by Hoffman's work \citep{Hoffman1995-uk} on ``eliciting knowledge from experts’’, we ask stakeholders to reflect on two aspects of domain expertise: 1) at a high level, they formalize a domain-specific workflow with its sub-tasks; and 2) at a low level, they externalize desired outcomes, values, and criteria of their decision-making process in each sub-task. The high level view enables participants to find potential design spaces where AI automation will be most useful, while the low level informs participants about their requirements, needs, and concrete parameters for developing practical AI solutions (e.g., input-outputs, training objectives, evaluation criteria, etc.).

Prior empirical studies in the domain provide a valuable starting point to learn about a stakeholder group and its activities. However, in a co-design environment, greater contextual information (e.g., use cases and user behavior) enables self-reflection for participants to better envision future new design opportunities \citep{Steen2013-ut}. Such ``tacit knowledge or latent needs'' are often difficult to express in words but can be documented via design activities, such as sketching, mapping, and modeling \citep{Visser2005-nt}. To support this, we create the \textit{Domain Specialty Canvas} (Figure \ref{fig:stakeholderspecialties}) based on the aforementioned low- and high-level aspects of domain expertise. This canvas helps fact-checkers to map their values and problems based on local work contexts. 

We also use think-aloud protocols to drive the mapping process. During the think-aloud protocol, we first let stakeholders describe different work cases. This helps inductively abstract a typical workflow. We then complement the protocol with semi-constructed interviews that ask stakeholders how they make decisions and solve problems in each task within the workflow. Finally, we ask stakeholders to reflect on the limitations of existing AI-assisted tools they have used in their work. 

\subsubsection{Playing with AI Probes (Step 2)} \label{subsec:AI_capabilities}

\begin{figure}[!t]
    \centering
    \includegraphics[width=0.9\textwidth]{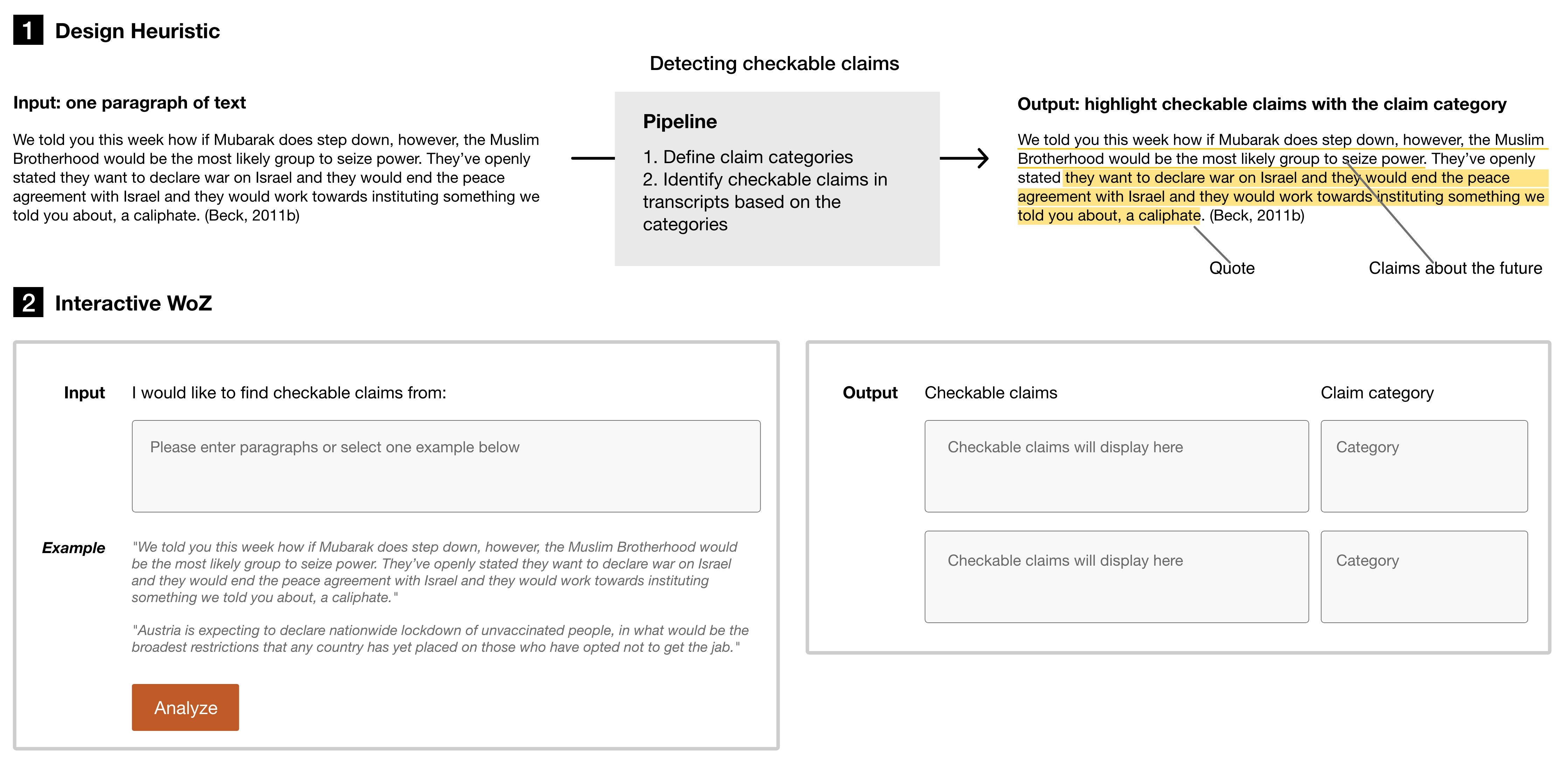}
    \caption{An \textit{AI Probe}: a half-baked AI idea consists of 1) design heuristic in the form of a textual diagram; and 2) abstracted NLP techniques in the form of an interactive Wizard-of-Oz simulation. This figure illustrates the first AI Probe (Detect checkable claims) presented to fact-checkers to help them learn the text classification technique.}
    \label{fig:designheuristic}
\end{figure}

In previous human-centered design (see Section \ref{subsec:co-designAI}), AI abstractions are often created to help non-AI experts to familiarize themselves with AI capabilities and limitations \citep{Subramonyam2021-yn, yang2019sketching, Jin2021-ik}. Such abstractions include ``visualizations, taxonomic vocabulary, or sensitizing concepts'' that reveal technological capabilities as simplified concepts \citep{yang2019sketching, Yildirim2023-ty}. In a similar vein, our second step leverages \textit{AI Probes} as AI abstractions (Figure \ref{fig:designheuristic}) to create a playground for stakeholders to have active conversations with AI researchers to become more familiar with AI concepts, behaviors, and possible errors.

As \citet{Van_Dijk2019-pm} discuss, it is difficult to map work activities to all possible techniques. In our matchmaking, AI Probes help facilitate this process. Specifically, each probe consists of two parts: 1) a visual design heuristic; and 2) an interactive WoZ (Wiz-of-Oz) demonstration. The visual design heuristic describes the basic elements of a half-baked AI idea as a pictorial diagram. It shows a standard NLP pipeline with inputs, outputs, and how inputs are processed to produce outputs. This diagram helps stakeholders understand AI from an end-user perspective without learning tricky technical details. The interactive WoZ demonstration is a design heuristic counterpart, which simulates an AI idea in a real-world context with simple functions and forms.

\begin{figure}[!t]
    \centering
    \includegraphics[width=0.8\textwidth]{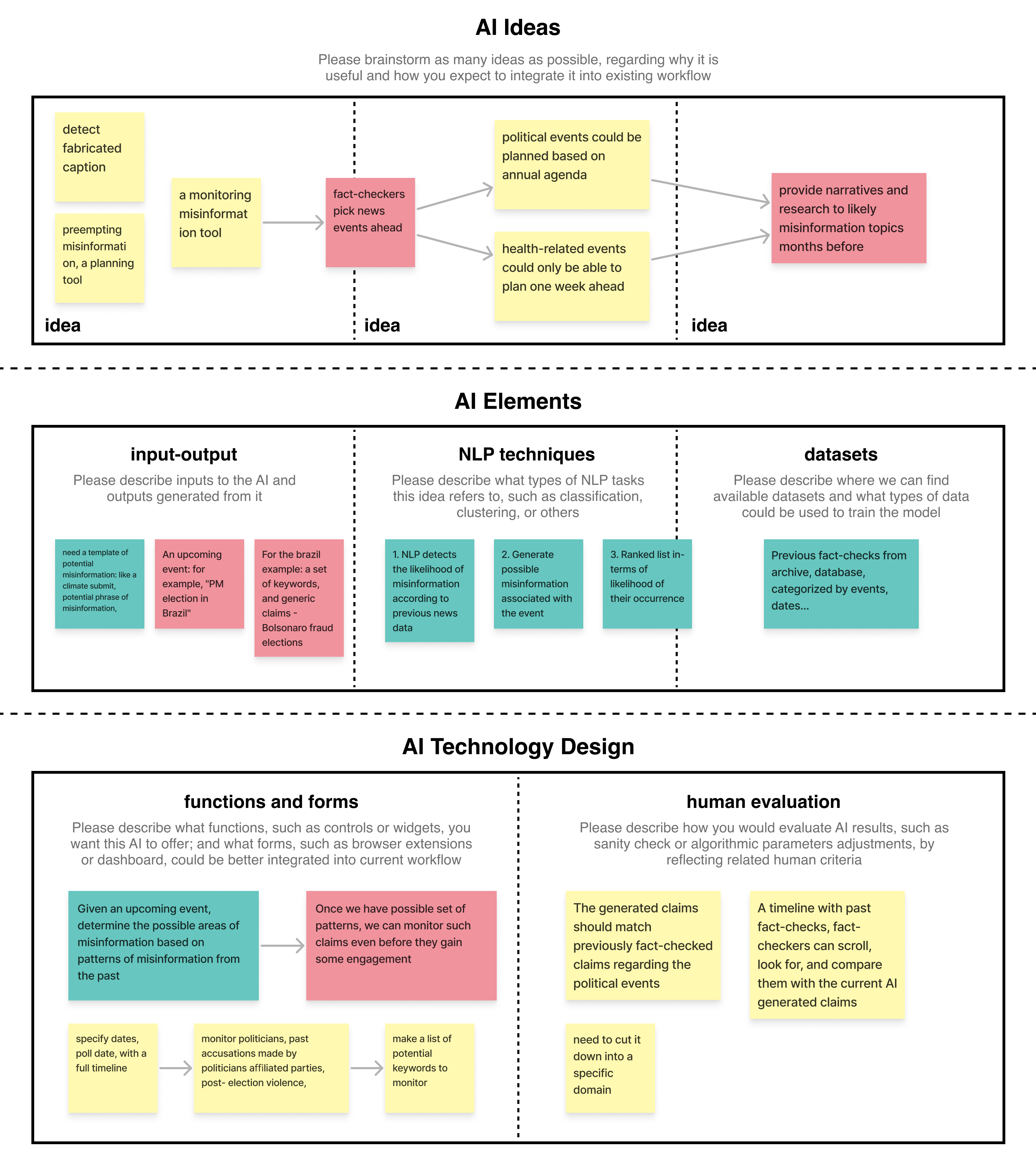}
    \caption{The \textit{Co-Design AI Canvas}: fact-checkers, designers, and AI researchers collaboratively brainstorm ideas by specifying different AI elements. This figure illustrates how Idea 1 (Section \ref{idea1}) is formulated. First, while fact-checkers were thinking aloud, designers helped them map out different seed ideas (i.e., yellow sticky notes). Continually, fact-checkers distilled the final one (i.e., red sticky notes and diagrams in the “AI ideas”). Later, AI researchers helped depict AI information (i.e., blue sticky notes in the “AI elements”). Meanwhile, fact-checkers provided concrete examples for these AI elements. Finally, all participants brainstormed “functions and form" and “human evaluation”.}
    \label{fig:feasibleAI}
\end{figure}

\subsubsection{Brainstorming AI Design Elements (Step 3)} \label{subsec:codesign_feasible_AI}

AI probes help facilitate participant capacity for brainstorming through three steps. First, the visual heuristics of these probes showcase different NLP techniques (e.g., text classification, clustering, and generations), so that participants become familiar with nuanced capabilities of NLP. Second, by playing with the interactive WoZ, participants gain insights into how different NLP techniques can be applied to assist their work. This interaction creates a link between the possibilities offered by AI and their specific needs, helping to develop a mental model for aligning ``technology affordance with work activity,'' as outlined by \citet{Bly1999-ar}. Finally, in the later brainstorming sessions, participants can reflectively assess the existing relationship between NLP capabilities and user activities originated from the probes, as well as suggest novel relationships where NLP techniques could be repurposed for different applications than originally intended.

To outline a feasible AI idea, previous human-centered design studies use {\em Canvas mapping} \citep{Axelsson2022-md, Subramonyam2021-yn, Zhou2020-ww} to carve out an AI lifecycle design. This includes: 1) creating a design brief with user personas; 2) specifying datasets and algorithms; and 3) designing prototypes with explainable user interfaces. Similar to these studies, we also create a \textit{Co-Design AI Canvas} to assist fact-checkers. Our canvas is more succinct, including only AI elements (AI needs, techniques, input-outputs, and datasets) and design elements (functions, forms, and evaluative techniques). Fact-checkers have been familiarized with these elements from the previous two activities, so they use this canvas to brainstorm ideas by mapping AI techniques they learned from the AI Probes to the domain expertise, values, and problems specified from the Domain Specialty Canvas. Fact-checkers fill in these essential elements with the assistance of facilitators (see more details in Figure \ref{fig:feasibleAI}). In this step, fact-checkers can either extend previous AI Probes into a comprehensive tool by combining different AI techniques or imagine completely new tools. AI researchers help introduce new AI techniques if stakeholders find techniques presented in the previous AI Probes cannot fully address their needs.

\subsection{Study Protocol} \label{subsec:protocol}

\begin{figure}[!t]
    \centering
    \includegraphics[width=\textwidth]{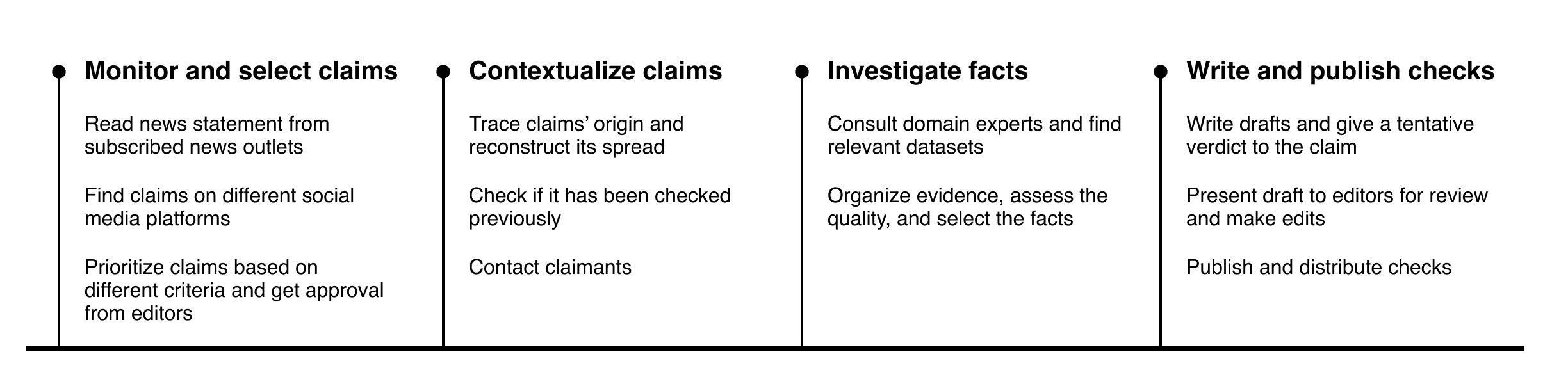}
    \caption{A fact-checking workflow synthesized from \citet{graves2017anatomy, micallef2022true} to guide participants to think-aloud their experiences.}
    \label{fig:factcheckingworkflow}
\end{figure}

In this section, we describe our workshop procedure and supplemental materials. They are developed to facilitate a two-phased workshop involving a recruited fact-checker, a designer, and an NLP researcher. The first phase was 1.5 hours long and focused on mapping fact-checker domain expertise (Step 1). The second phase was 2 hours long and included playing with AI Probes (Step 2) and brainstorming NLP-based AI ideas (Step 3). We used Miro\footnote{\url{https://miro.com/}} to present co-design materials and document participant feedback.

\subsubsection{Workshop Procedure}

\begin{table}
\renewcommand{\arraystretch}{1.2}
  \resizebox{\textwidth}{!}{%
  \begin{tabular}{l>{\raggedright}p{3cm}>{\raggedright}p{2.5cm}>{\raggedright}p{4.5cm}>{\raggedright\arraybackslash}p{3cm}}
    \toprule
    No.&AI Probes&AI behaviors&Details&Related NLP research\\
    \midrule
    \multicolumn{5}{l}{%
    \textit{NLP helps to identify fact-checking related qualities}} \\

    1&Detect checkable claims&Text Classification&Classify sentences into claims categories, including personal experiences, public opinions, quantitative/numeric claims, causation and correlation&claim detection\citep{Konstantinovskiy2021-od}\\
    2&Detect social media checkworthy claims&Text Classification&Identify checkworthy claims based on posts virality and highly-emotional content&virality and susceptibility\citep{Hoang2012-ak}, ideology detection\citep{Xiao2020-ew}\\
    3&Review the argumentation quality of a fact-check* &Text Classification&Evaluate the readability, coherence, and persuasiveness of writing&detect attackable sentence\citep{Jo2020-qv}\\
    \hline
    \multicolumn{5}{l}{%
    \textit{NLP helps to prepare claims in a format that is convenient for fact-checking}} \\
    
    4&Prepare stand-alone claims for fact-checking* &Text Classification + Text Generation&Add contextual information to a claim from previous paragraphs&decontextualization\citep{Choi2021-bm}\\
    5&Identify ambiguous terms in a claim* &Text Classification + Question Generation&Generate sub-questions from ambiguous terms of a claim for later investigation&generate fact-check brief\citep{Fan2020-sj,Chen2022-yz}, political ambiguity\citep{Chapp2019-lo}\\
    \hline
    
    \multicolumn{5}{l}{%
    \textit{NLP helps to extract additional data in an accessible, effective form to help fact-checking}} \\
    
    6&Summarize claim-related articles* &Text Summarization&Summarize a claim's related articles into bullet points&extractive and abstractive fact-check explanations\citep{Kazemi2021-qn}\\
    7&Retrieve quantitative related information&Text Retrieval +
    Text Summarization&Retrieve and format quantitative information consistently into table from the search results&unstructured data for analytical queries\citep{Li2021-ff}\\
    \hline
    
    \multicolumn{5}{l}{%
    \textit{NLP helps to determine veracity of claims for just in time claim-checking}} \\
    
    8&Robocheck by evidence retrieval&Text Retrieval + 
    Text Inference&Auto-verify the true and false of a claim by retrieving claim related information&detect fake news from crowd\citep{Nguyen2018believeornot}\\
  \bottomrule
\end{tabular}
}
\vspace{0.2cm}
    \caption{Eight \textit{AI Probes} were prepared to help stakeholders understand what existing NLP could offer to assist their work; ideas marked with * incorporate relatively new NLP techniques with potential value for fact-checking.}
  \label{tab:nlpideas}
\end{table}

For Step 1, we asked fact-checkers to bring specific claims they checked in the past, including: 1) a claim they found particularly difficult to check; 2) one whose veracity surprised them after checking; 3) their most recent check; and 4) a claim they found particularly interesting. After reflecting on these four claims, participants walked us through their typical workflow, selecting one of these claims as a working example. To facilitate this process, we prepared a fact-checking diagram (Figure \ref{fig:factcheckingworkflow}) as a reference. This served as a useful memory aid to remind fact-checkers of the various sub-tasks they may neglect to mention. 

For Step 2, we created eight specific \textit{AI Probes}, which map existing state-of-the-art NLP techniques to different fact-checking tasks (Table \ref{tab:nlpideas}). As discussed in Section \ref{subsec:AI_capabilities}, past NLP research proposed for fact-checking considered some potential ways NLP could support professional fact-checkers. However, it is important that fact-checkers verify these scenarios, as well as brainstorm themselves new scenarios or how NLP techniques might support other work activities. We organized visual design heuristics for eight probes onto a one-page ``fact sheet'' and encouraged fact-checkers to read it before the second workshop. We also asked them which NLP techniques they were most interested in, then had them experiment with the eight interactive WoZ demos, one-by-one. 

For Step 3, we created a prompt protocol for each participant. The protocol included questions framed around a typical fact-checker workflow (documented in the \textit{Domain Specialty Canvas}). For example: “Based on challenges identified in your workflow, do you see any opportunities to apply the AI techniques we have discussed?” Such questions oriented fact-checkers toward bridging AI techniques with their respective needs and to spur ideation. Fact-checkers then worked alongside facilitators to brainstorm ideas by filling in a set of AI design elements on the \textit{Co-Design AI Canvas}.

\subsubsection{AI Probe Design and Selection}
We began by reviewing survey papers on automated fact-checking. This involved analyzing relevant NLP techniques adopted for automated fact-checking and existing computational fact-checking tools highlighted by nonprofit research organizations (e.g, the RAND Corporation\footnote{\url{https://www.rand.org/research/projects/truth-decay/fighting-disinformation/search.html}} and Credibility Coalition\footnote{\url{https://credibilitycoalition.org/credcatalog/}}) that contain resources for fact-checking. For the workshop, we choose specific probes (1, 2, 4, and 7) based on fact-checker unmet needs identified in recent empirical work (e.g., monitoring and prioritizing claims \citep{micallef2022true, Procter2023-rr} and augmenting ambiguous claims \citep{Singh2021-fj}). Research has reported that these needs are recognized but not adequately addressed by current tools \citep{ArnoldPh:online}. We also designed probes (3, 5, 6, and 8) based on other use cases proposed by NLP researchers. While these methods have yet to be integrated into practical tools, we included them as AI probes to gauge fact-checker reactions and assessment of potential utility.

% that while not yet supported by existing tools, NLP capabilities have been demonstrated bin research (see Table \ref{tab:nlp-based fact-checking solutions}).}

% 
% 

To develop probes to realistically simulate NLP capabilities for participant use, we employed a variety of methods. We utilized readily available APIs to summarize claim-related articles for Probe 6, as well as automatically verifying new claims for Probe 8. We also constructed basic Naive Bayes models for sorting claims into categories of {\em checkability} (Probe 1) and check-worthiness (Probe 2). In cases where we could not easily generate outputs using advanced models (i.e., Probes 3, 4, 5, 7), a mix of accurate and inaccurate results was produced based on the test examples presented in NLP research (listed in Table \ref{tab:nlpideas}).

\subsection{Recruitment} \label{subsec:recruitment}

%\begin{sidewaystable}
%\begin{center}
\begin{table}[ht]
% \centering
\resizebox{\textwidth}{!}{%
\begin{tabular}{@{}llllllllllll@{}}
\toprule
\textbf{No.}& \textbf{\begin{tabular}[c]{@{}l@{}}1st\end{tabular}} & \textbf{\begin{tabular}[c]{@{}l@{}}2nd\end{tabular}} & \textbf{\begin{tabular}[c]{@{}l@{}}Age \\ Group\end{tabular}} & \textbf{Gender} & \textbf{Ethnicity}                                                          & \textbf{Role}        & \textbf{\begin{tabular}[c]{@{}l@{}}Organizational\\ Context\end{tabular}} & \textbf{\begin{tabular}[c]{@{}l@{}}Languages \\ Fact-Checked\end{tabular}} & \textbf{\begin{tabular}[c]{@{}l@{}}Years of \\ Fact-\\ Checking\\ Experience\end{tabular}} & \textbf{Country} & \textbf{Region} \\ \midrule
1         & $\checkmark$                                     & $\checkmark$                                      & 26-35                                                         & Female          & White                                                                       & Fact-checker         & Ambiguous                                                                 & English                                                                    & 1                                                                                          & USA              & North America   \\ 
2         & $\checkmark$                                     & $\checkmark$                                      & 36-45                                                         & Male            & Black                                                                       & Fact-checker         & Independent                                                               & English                                                                    & 14                                                                                         & USA              & North America   \\ 
3         & $\checkmark$                                     &                                                                   & 46-55                                                         & Female          & White                                                                       & Editor               & Independent                                                               & English                                                                    & 15                                                                                         & USA              & North America   \\ 
4         & $\checkmark$                                     & $\checkmark$                                      & 26-35                                                         & Female          & White                                                                       & Fact-checker         & Independent                                                               & \begin{tabular}[c]{@{}l@{}}English \& \\ Afrikaans\end{tabular}            & 0.5                                                                                        & South Africa     & Africa          \\ 
5         & $\checkmark$                                     & $\checkmark$                                      & 46-55                                                         & Male            & Indian                                                                      & Editor & Independent                                                               & \begin{tabular}[c]{@{}l@{}}English \& \\ Nepali\end{tabular}               & 2.5                                                                                        & Nepal            & Asia            \\ 
6         & $\checkmark$                                     & $\checkmark$                                      & 26-35                                                         & Female          & Indian                                                                      & Fact-checker         & Independent                                                               & English                                                                    & 3                                                                                          & South Africa     & Africa          \\ 
7         & $\checkmark$                                     &                                                                   & 26-35                                                         & Female          & No response                                                                 & Editor               & Independent                                                               & Turkish                                                                    & 5                                                                                          & Turkey           & Europe          \\ 
8         & $\checkmark$                                     & $\checkmark$                                      & 18-25                                                         & Male            & White                                                                       & Fact-checker         & Independent                                                               & \begin{tabular}[c]{@{}l@{}}English \& \\ Afrikaans\end{tabular}            & 2                                                                                          & South Africa     & Africa          \\ 
9         & $\checkmark$                                     & $\checkmark$                                      & 46-55                                                         & Male            & \begin{tabular}[c]{@{}l@{}}Pākehā/European-\\ Australian/White\end{tabular} & Fact-checker         & Media                                                                     & English                                                                    & 2                                                                                          & New Zealand      & Oceania         \\ 
10        & $\checkmark$                                     &                                                                   & 26-35                                                         & Female          & Aryan                                                                       & Fact-checker         & Independent                                                               & Nepali                                                                     & 5                                                                                          & Nepal            & Asia            \\ 
11        & $\checkmark$                                     & $\checkmark$                                      & 18-25                                                         & Male            & Hindu, Bhramin                                                              & Fact-checker         & Independent                                                               & \begin{tabular}[c]{@{}l@{}}Nepali \& \\ English\end{tabular}               & 0.3                                                                                        & Nepal            & Asia            \\ 
12        & $\checkmark$                                     & $\checkmark$                                      & 18-25                                                         & Male            & Caucasian                                                                   & Fact-checker         & Media                                                                     & English                                                                    & 2                                                                                          & USA              & North America   \\ 
13        & $\checkmark$                                     & $\checkmark$                                      & 26-35                                                         & Female          & Black African                                                               & Fact-checker         & Independent                                                               & English                                                                    & 2                                                                                          & South Africa     & Africa          \\ 
14        & $\checkmark$                                     & $\checkmark$                                      & 18-25                                                         & Female          & White                                                                       & Fact-checker         & Media                                                                     & English                                                                    & 1.5                                                                                        & USA              & North America   \\ 
15        & $\checkmark$                                     & $\checkmark$                                      & 26-35                                                         & Female          & \begin{tabular}[c]{@{}l@{}}Pākehā/European-\\ Australian/White\end{tabular} & Fact-checker         & Media                                                                     & English                                                                    & 0.5                                                                                        & New Zealand      & Oceania         \\ 
16        & $\checkmark$                                     & $\checkmark$                                      & 18-35                                                         & Male            & \begin{tabular}[c]{@{}l@{}}Prefer not to\\ Answer\end{tabular}              & Fact-checker         & Independent                                                               & \begin{tabular}[c]{@{}l@{}}English \& \\ Filipino\end{tabular}             & 3                                                                                          & Philippines      & Asia            \\ 
17        & $\checkmark$                                     & $\checkmark$                                      & 26-35                                                         & Male            & Hispanic                                                                    & Fact-checker         & Media                                                                     & \begin{tabular}[c]{@{}l@{}}English \& \\ Spanish\end{tabular}              & 2                                                                                          & Mexico           & Latin America   \\ 
18        & $\checkmark$                                     & $\checkmark$                                      & 26-35                                                         & Male            & White                                                                       & Fact-checker         & Independent                                                               & English                                                                    & 7                                                                                          & USA              & North America   \\ 
19        & $\checkmark$                                     & $\checkmark$                                      & 56-65                                                         & Male            & White                                                                       & Editor               & Independent                                                               & English                                                                    & 7                                                                                          & USA              & North America   \\ 
20        & $\checkmark$                                     & $\checkmark$                                      & 26-35                                                         & Male            & White British                                                               & Fact-checker         & Media                                                                     & English                                                                    & 3.5                                                                                        & England          & Europe          \\ 
21        & $\checkmark$                                     & $\checkmark$                                      & 26-35                                                         & Male            & Asian                                                                       & Fact-checker         & Media                                                                     & English                                                                    & 1                                                                                          & India            & Asia            \\ 
22        & $\checkmark$                                     & $\checkmark$                                      & 56-65                                                         & Female          & Prefer not to answer                                                        & Fact-checker         & Independent                                                               & English                                                                    & 1                                                                                          & USA              & North America   \\ \bottomrule
\end{tabular}%
}
\vspace{0.2cm}
\caption{Participant Demographics and Work History}
\label{tab:participant}
\end{table}
%\end{center}
%\end{sidewaystable}

We recruited (\textit{n} = 22) professional fact-checkers, including 5 present or former editors and 17 full-time fact-checkers (Table \ref{tab:participant}). All editors had previous fact-checking experience and often guide their organizational fact-checkers. All participants participated in the first workshop session, and 20 participated in the second session. Our first workshop session lasted 1.5 hours, and the second lasted 2 hours. Each participant was compensated with Amazon Gift Cards or other redeemable options available in their areas. 

Participants (11 women and 11 men) were drawn from 9 countries, including the USA, South Africa, Nepal, Turkey, New Zealand, Philippines, Mexico, England, and India. We intentionally sampled participants from diverse countries, fact-checking topics, organizations, ages, genders, etc. Among the participants, 11 (52.4\%) identified as White, 5 (23.8\%) as Asian, 2 (9.5\%) as Black, 1 (4.8\%) as Hispanic, and 2 (9.5\%) did not self-identify. 5 (23.8\%) identified as being in the 18-25 years old age range, 11 (52.4\%) in the 26-35 range, 1 (4.8\%) in the 36-45 range, 3 (14.3\%) being between 46-55, and 1 (4.8\%) in the 56-65 age range. Their fact-checking experience ranged from 6 months to 15 years, with an average of 3.8 years. 18 were from independent fact-checking organizations, while the remaining 4 were from news media outlets. Upon asking fact-checkers about their familiarity with AI, most of our participants had little knowledge of computational algorithms and programming, yet most had used different computational tools to assist their work. These include: 1) Open-source Intelligence (OSINT) tools, such as the InVid; 2) specific claim monitoring tools, such as CrowdTangle, TweetDeck, and Full Fact Alpha; 3) off-the-shelf search engines, including Google and Bing; and 4) databases, such as the Internet Archive and LexisNexis.

\subsection{Analysis} \label{subsec:analysis}
We conducted all co-design sessions with recruited participants on Miro and Zoom and then transcribed the recordings for analysis. Building upon previous empirical studies \citep{graves2017anatomy, micallef2022true}, we generated a set of codes for existing fact-checking practices to analyze transcripts from Step (1). Other codes were generated via a bottom-up thematic analysis approach based on how fact-checkers played with AI Probes in Step (2) and what ideas they brainstormed in Step (3) \citep{patton2014qualitative}. The qualitative coding was initially conducted by a single primary researcher and then reviewed by a second researcher. During the open coding, the primary coder examined transcripts on a sentence-by-sentence basis, examining whether any findings could be classified by existing literature or whether a new code needed to be created. All of these codes are divided into three parts: fact-checking criteria, stakeholder design experience, and co-designed ideas. We aggregate the fact-checking criteria into tables in Appendix \ref{appendix:criteria-1} and \ref{appendix:criteria-2}. Our Results (Section \ref{sec:results}) primarily focus on AI ideas co-designed with fact-checkers. Our Discussion (Section \ref{sec:discussion}) then summarizes fact-checker design experience as evidence to evaluate the utility of our co-design method. \added{As our participants represent organizations in multiple countries, we also analyzed the potential influence of regions and organizational interests on participant needs and co-created ideas. Given that we have a few participants representing these different groups, we report them in Appendix \ref{appendix:criteria-3} as exploratory findings.}

\subsection{Positionality Statement} \label{subsec: positionality}

Our research team has members with diverse backgrounds, including HCI, design, NLP, information retrieval, crowdsourcing, and AI ethics. The main workshop facilitators include one UX professional with years of industry experience and another UX researcher who has received academic training in HCI. The two NLP researchers who participated in the second workshop activity have expertise in explainable AI and NLP applications for misinformation. The interdisciplinarity of our team allowed us to take a human-centered approach to develop {\em Matchmaking for AI} and our co-design study. We take a constructive stance, believing that NLP has the potential to assist in fact-checking. 

As both designers and NLP researchers are also co-authors of this paper, we recognize that our perspectives and roles naturally influenced the research process. Bearing this reflexivity in mind, we sought to minimize researcher-induced biases. First, we invited an expert with over 15 years of experience in fact-checking research and misinformation reporting to refine our workshop scripts and study procedures. Second, during the matchmaking, we encouraged participants to brainstorm ideas by pinpointing problems reported in the \textit{Domain Specialty Canvas}, ensuring that ideas originated directly from participants.
%Lastly, as described in the Section \ref{subsec:analysis}, we formed two teams, each including one designer and one NLP engineer, who separately participated in the workshop and carried out the thematic coding via a peer-review process. 
We further discuss the potential limitations of our study in Section \ref{subsec:limitations}.

\section{Results} \label{sec:results}

In this section, we describe the ideas generated from our co-design workshops that brought together AI experts, designers, and professional fact-checkers. Given our definition of ``workable ideas'' (Section \ref{subsec:matchmaking}), for each idea, we provide details on stakeholder needs, gaps in existing tools and related work in AI research, desired tool features, and potential technical and research challenges. These details are synthesized from participant notes on the \textit{Domain Specialty Canvas} and the \textit{Co-design AI Canvas}, along with their reflections on the \textit{AI Probes}. 
% Unlike a speculative co-design approach that asks end-users to brainstorm solutions without considering the AI capabilities upfront, \textit{Matchmaking for AI} helps generate ``workable'' ideas that are both grounded in real participant needs and informed by an understanding of realistic AI capabilities. 
% We identify these realistic ideas as \textit{workable}. In the context of our study, an idea is workable if it is grounded in existing AI capabilities. 
% Note that grounded ideas do not necessarily mean they can be implemented directly with off-the-shelf models.
While there are parallels to the workable ideas in related AI research (described in Table \ref{tab:nlp-based fact-checking solutions}), 
% \ml{cite} 
most of these ideas require building datasets specific to the tasks and additional domain-specific modeling work for implementation. Moreover, we identify if there is a need for more research in AI capabilities to implement them, in turn providing a road map for AI researchers interested in fact-checking. 
% Compared to existing co-design literature that usually provides only a conceptual discussion of the stakeholder needs, for each workable idea, we provide more details on stakeholder needs, gaps in existing tools and related work in AI research, desired tool features, and potential technical and research challenges. 
% Some ideas presented here require more NLP research in task design, dataset creation, and model development. 
We expect further technical and design work will be needed to develop, implement, and deploy these ideas in practice.

% a) need assessment, b) specific gaps identified from the existing tools and related prior work in the literature, c) tool features and interactive functions desired by our participants, d) mapping to existing AI techniques, and e) potential technical challenges. }

% Although our co-design objective is to develop ``workable'' ideas that provide concrete design suggestions and technical details, we acknowledge that the concept of what constitutes a ``workable'' idea is somewhat subjective. Thus, during our analysis, we transform the insights gathered from the co-design AI canvas and discussions between participants and NLP researchers into a structured written format. In this section, each idea is written in such a format to depict its feasibility. Specifically, each idea begins with a need assessment to highlight its importance by presenting a brief summary of related prior work (either from the academic or industrial practice) and the current gap identified by our participants. Following this, we present design details, including tool features and interactive functions conceptualized by participants along with useful NLP techniques. These techniques were a consensus between participants and NLP researchers, developed through their collective reflection on the AI Probes. Finally, we present the potential challenges faced by implementing these ideas in practice, discussing both the difficulties users may face in adopting them and the infrastructural obstacles encountered during AI development.}

Table \ref{tab:nlp-based fact-checking solutions} summarizes the 11 co-designed ideas, which span the fact-checking lifecycle from claim monitoring, selection, and investigation to the final fact-check writing stage. Compared with the dominant cases of using NLP techniques for fact-checking --- detecting and ranking claim check-worthiness, retrieving previously fact-checked claims and evidence, and auto-verifying claims \citep{nakov2021automated, Das2023-tq} --- we find other interesting use cases, such as forecasting disinformation (Idea 1, Section \ref{idea1}), contextualizing ambiguous claims (Idea 4, Section \ref{idea4}), human-AI teaming for generating fact-check briefs (Idea 7, Section \ref{idea7}) and co-writing fact-check reports (Idea 10, Section \ref{idea10}). We categorize these use cases as design ideas and highlight their novelty based on: 1) whether this design idea is completely new to fact-checking; or 2) whether our participants uncovered new needs not identified in prior work. In the following sections, we present ideas that either reflect new design or address unmet needs. We also synthesize ideas from participants that appear significant even if they are not particularly novel. We include these in Appendix \ref{appendix:existing-needs-designs}. \added{In Appendix \ref{appendix:criteria-3}, we present an exploratory analysis on how different regions and organizational interests influence participant needs and co-created ideas.}

% Please add the following required packages to your document preamble:
% \usepackage{booktabs}
% \usepackage{multirow}
% \usepackage{bbding}
\begin{table}[ht]
    \resizebox{\textwidth}{!}{%
    \begin{tabular}{@{}p{2cm}l>{\raggedright}p{5cm}>{\raggedright}p{1.1 cm}>{\raggedright}p{1cm}>{\raggedright\arraybackslash}p{2cm}@{}} \toprule
        Fact-checking stages & No. & Design ideas & New idea & New need & Related AI research \\ \midrule
            Forecasting claims & $1$ & Forecasting Disinformation & $\checkmark$ & $\checkmark$ & \cite{Huang2022-qh, Pan2023-me}\\ \midrule
            \multirow{2}{2cm}{Monitoring claims} & $2$ & Identifying Broader Disinformation Narratives &  & $\checkmark$ & \cite{Jiang2021-ef, Hardalov2021-zm, Arakelyan2023-ju}\\ \cmidrule(r){2-6} 
            & $3$ & Dynamic Credibility Monitoring of Social Media Users & $\checkmark$ & $\checkmark$ & \cite{zhou2020survey, Nguyen2018believeornot, Giachanou2019-qf, Zhi2017-wv, Rashkin2017-ct}\\ \cmidrule(r){1-6} 
            \multirow{3}{2cm}{Selecting claims} & $4$ & Finding and Providing Context for Ambiguous Claims & $\checkmark$ &  & \cite{Choi2021-bm, Chen2022-yz}\\ \cmidrule(r){2-6} 
            & $5$ & Personalized Claim Filtering and Selection &  & $\checkmark$ & \cite{alam2020fighting, Hassan2017TowardAF, vasileva2019takes, Nakov2022TheCC, alam2021fighting, Gupta2023-jx, Kartal2020-hp, Hardalov2021-zm, Arakelyan2023-ju}\\ \cmidrule(r){2-6} 
            & $6$ & Personal Bias Warning System for Claim Selection & $\checkmark$ & $\checkmark$ & \cite{Hardalov2021-zm, Arakelyan2023-ju}\\ \cmidrule(r){1-6}
            \multirow{3}{2cm}{Investigating facts} & $7$ & Human-AI Teaming for Fast Generation of Fact-check Briefs & $\checkmark$ & $\checkmark$ & \cite{Fan2020-sj, Chen2022-yz, Nguyen2020-we, Chen2023-fu}\\ \cmidrule(r){2-6} 
            & $8$ & Identifying Official Databases and Formulating Queries for Verifying Quantitative Claims & & $\checkmark$ & \cite{Saeed2021-ff, Karagiannis2020ScrutinizerAM, Balalau2022-wu}\\ \cmidrule(r){1-6}
            \multirow{3}{2cm}{Writing fact-checks} & $9$ & Understanding Reader Engagement for Fact-checking Reports & $\checkmark$ & & \cite{Zhang2021-mh}\\ \cmidrule(r){2-6} 
            & $10$ & AI Assistance in Writing Fact-checking Reports & $\checkmark$ & $\checkmark$ & \cite{Chen2023-fu}\\ \cmidrule(r){2-6} 
            & $11$ & AI Assistance in Critiquing and Editing Fact-checking Reports  & & $\checkmark$ & \cite{Vo2019-mr}\\ \bottomrule
	\end{tabular}
	}
    \vspace{0cm}
    \caption{Matchmaking-for-AI outcomes across design ideas, AI research, tools, and resources. This table describes the outcome of our co-design process. Our process led to identification of 9 novel needs for fact-checkers and 7 new ideas that can inform future tools for fact-checking. Additionally, we provide pointers to the literature related to the design ideas that can be used by developers as a starting point.} \label{tab:nlp-based fact-checking solutions}
\end{table}

\subsection{\textbf{Forecasting Disinformation}} \label{idea1}

Disinformation and misinformation frequently reoccur. A simple example is when old fake news is ``recycled'' (often verbatim or with only minor tweaks) and re-distributed anew. Slightly more sophisticated is when certain types of news events (e.g., an election or a pandemic) are routinely followed by regular patterns of disinformation (e.g., \textit{``the presidential candidate is associated with <insert any crime>''}) \cite{swire2022backfire}. Participants noted that before major events like presidential elections, they often release ``fact sheets'' or ``misinformation fact-check guides'' to the public to help ``inoculate'' them against potential fake news exposure. However, this process typically requires extensive analysis of social media engagement, such as identifying what false news the public is likely to believe as true (P4, 19) or relies on fact-checker intuition to predict potential issues (P4). Thus, participants wanted to understand if and how AI could systematically assist in pre-bunking. Specifically, they wanted AI to help identify which facts are more crucial to report on and prevent the possible recurrence of similar false claims: ``If three months in advance I'm already aware of misinformation topics that would spread, I will have my coverage ready to counter those narratives... [so that] these misinformation claims don't actually gain traction'' (P21).

Generating misinformation has mainly been seen by AI researchers as a method to mimic possible misuse of LLMs \citep{Pan2023-me} or to evaluate the effectiveness of new misinformation detection tools \citep{Huang2022-qh}. In contrast, our study participants have redirected the use of a generative model toward a more beneficial use scenario, i.e., forecasting potential fake news for recurring news events to be better prepared for fact-checking potential misinformation.

During the matchmaking Step (3), participants (P4, 21) and NLP researchers brainstormed the use of generative models to create such potential misinformation ahead of an upcoming event, e.g., given context of the event type. Models might generate a set of possible fake claims or a ranking ordered by their predicted probability of occurrence based on historical data. For example, to forecast possible fake news claimed by candidates for an election, P21 said the AI tool should be able to ``identify possible set of [discourse] patterns or narratives on [historical] fake results, rumors, and previously fact-checked claims.'' Additionally, it would be beneficial to provide data insights, such as summarizing how voters from various geographical regions have significantly engaged with these claims on social media (connected to Idea 9, Section \ref{idea9}). Such historical evidence and insights could bolster their trust in use of AI: ``I can trace it back to understand: Are there any facts on these things [AI predictions]?'' Regarding specific tool features, participants imagined that a timeline would be particularly useful, allowing them to scroll, look for, and compare previous claims, fact-checks, and historical data with the AI-generated claims. 

While language models have been used to generate variants of existing claims (e.g., counter-factual claims \citep{saakyan2021covid}), predicting future fake news would be significantly different for model training and evaluation. Researchers could curate new datasets that pair examples of past news events with associated fake news related to those events. Participants also suggested a way to evaluate the prebunking models before trusting model outcomes in practical scenarios. For example, P21 suggested creating a test set by putting aside claims from a known source, training a model on older claims, and testing if the model could produce unseen claims in the test set. Even when the model is not 100\% accurate, the participants deemed the prebunking model helpful to guide their intuition in predicting what claims might occur in the future. In the context of misinformation in political speech, P21 said, ``it still helps you understand the candidate’s background, chase information on this topic ...''. Participants thought it could be helpful in trying to anticipate all possible misinformation that might occur. 

% and particularly because I was aware of the maximum number of outputs [what all possible fake news might re-occur].''} 

% \added{However, the practical usefulness of this AI tool might vary greatly, depending largely on the event topics and datasets used. P21 explained that because political claims are often vague, if AI-generated claims are similarly nonspecific without targeted information of what is false, it might not be effectively useful for pre-bunking. Thus, to enable AI to generate both precise and probable claims, the dataset needs to be meticulously curated, focusing on claims and fact-checks labeled as strictly false.} 

\subsection{\textbf{Identifying Broader Disinformation Narratives }} \label{idea2}

Individual claims often reflect larger narrative patterns in thematically connected claims. For example, a theme of \textit{distrust in technology creators and providers} might connect claims such as: a) Bill Gates is using the COVID-19 vaccine to implant microchips, b) self-driving cars self-destructing, and c) Twitter uses improper means to censor content (P7). Several participants (P1, 7, 19, 21, 22) refer these narratives as ``conspiracy theories'' or ``hoax'' that can easily provoke social media users for sharing. Recognizing such narratives are important, assisting fact-checkers to: 1) monitor disinformation by understanding more on what to search for (see Idea 5, Section \ref{idea5}), and 2) develop counter-strategies against these narratives when writing fact-checks.

Communication studies \citep{Douglas2021-gf, Douglas2019-tk} suggest that people often believe and share disinformation narratives for psychological or social needs, such as the desire of curiosity, avoiding uncertainty, or group belonging. Our participants (P7, 21) note that by recognizing these narratives, they can ``understand motivations [people] share and how disinformation affects different groups'', and informs them to write ``about the truth to convince people more easily'’ (P7). Fact-checking reports informed by the broader narrative might help alleviate feelings of collective insecurity or discomfort caused by the disinformation. However, P7 points out that existing tools do not provide insights into broader narratives or connections between different ones; they have to manually piece together potential narratives, or perform a manual retrospective analysis from previously fact-checked claims.

Participants were interested in whether NLP methods could help identify such larger narratives. Specifically, two settings were envisioned: 1) given a set of claims, identify the broader narrative; and 2) given a narrative statement, generate claims to fit the narrative. 

For the first, as with the previous forecasting idea (Section \ref{idea1}), generating such claims could help fact-checkers prepare for anticipatory pre-bunking. P7 gave an example of generation from a narrative, ``vaccines are bad for people'', with possible false claims, such as ``vaccines are a game of US and China'' or ``Bill Gates called for the withdrawal of vaccines.'' NLP techniques such as topic modeling or unsupervised clustering could help group claims together to identify claims that have similar narratives \citep{Arakelyan2023-ju, Jiang2021-ef}. Furthermore, generative models built to provide natural language descriptions of such clusters could help describe the narratives. 

For the second setting, a generative language model could be trained to produce example claims given a narrative. For example, a state-of-the-art generative model (such as GPT-4\footnote{\url{https://openai.com/research/gpt-4}}) might be applied in a zero-shot or few-shot manner \citep{wei2021finetuned}. However, because generative AI providers are increasingly wrapping their service offerings with ``guardrails'' to prevent adversarial use (such as generating disinformation, as envisioned here), it may be necessary to use an open-source model or obtain privileged API access without the guardrails. In addition, evaluating the quality of generated claims would require human review and/or an annotated dataset. 

\subsection{\textbf{Dynamic Credibility Monitoring of Social Media Users}} \label{idea3}

As an example of the 80/20 rule, a disproportionately large amount of misinformation tends to be spread by a relatively small handful of online sources (e.g., social media users). Participants reported that they often manually curate and monitor content from known, repeat offenders. This led to the suggestion (P5, 19-21) of developing a ``credibility checker'' tool that could automatically estimate and continually update the predicted credibility of such users (e.g., based on the content they post, their social network of followers and who they follow, etc.). To assess the authenticity of social media profiles, related prior work has commonly employed tools that are used to identify social media bots, such as the Bot sentinel or Botometer \citep{Rauchfleisch2020-tz, hays-webconf23}. In our study, participants advocated for the value of long-term tracking of personal credibility of real individuals. 

Note that the {\em Global Disinformation Index} (GDI)\footnote{\url{https://www.disinformationindex.org/}} \citep{srinivasan2019-nb} already provides a ``neutral, independent, transparent index of a website’s risk of disinforming readers.'' Similarly, existing claim detection methods already pay attention to individual posts or known malicious sources at an organizational level \cite{zhou2020survey}. As participants (P5, 14, 20) noted based on the 1st and 2nd AI probes, claim detection techniques having individual credibility ratings could be used to achieve this. The key technical challenge would be modeling and estimation of user credibility, and the frequency of ingesting new data and updating model estimates for each user. 

Political and ethical considerations around such credibility monitoring were also discussed. For example, GDI has faced intense criticism and litigation in response to its domain-level source credibility ratings. There are also risks of harm from assigning low credibility scores due to AI detection errors, especially if those scores became public. Additionally, there are questions of how easily a reformed user could overcome a low credibility score given a past history of bad behavior. During the matchmaking Step (3), however, participants did not flag any strong ethical concern towards building such a tool, as many already monitor potentially problematic sources of misinformation internally. P14's organization has a long history of monitoring and sharing politician records regarding sharing false claims. Similarly, P19 stated that they released information about social media users who are probable sources of misinformation. Additionally, fact-checking programs from Twitter (now X) and Facebook were frequently mentioned by participants as community efforts committed to overseeing misinformation spreaders. 
% However, participants also noted that credibility monitoring tools must not bias the fact-checks on new claims.
% and only help to alleviate ethical concerns.}
%We think it remains vital to maintain the credibility score records private if such a technology is implemented.  while ensuring that the fact-checks are reported transparently, with clear justifications for the origin of the claims.}

\subsection{\textbf{Finding and Providing Context for Ambiguous Claims}}\label{idea4} 

Textual claims on social media often lack sufficient context to be clearly understood \citep{Singh2021-fj}. 
Claim checking such texts often requires additional disambiguation and world knowledge. 
% the practical adoption of these techniques in fact-checking has not been fully discussed.}
Participants raised the idea of using AI to augment a claim with contextual information so that the claim's terms, concepts, and/or arguments are less ambiguous. This stemmed in part from our AI probe using stand-alone claims, leading some participants to add information about what needed to be disambiguated. P5 explained that ``what misinformation does is it removes the context''. For example, in the claim that ``Midtown crime is up by 30\% the last quarter'' (taken from \cite{Singh2021-fj}), ``midtown'' could refer to Midtown Manhattan in New York City, or several cities in the US state of Florida. P1 voiced that the contextual information could ``help us determine whether or not that there was a basis for the claim and if that's something we needed to do a deeper dive into''. This contextual information could include background knowledge (e.g., time and location - P5) and cultural context (e.g., who is the targeted audience and who will be interested in the claim - P5, 11, 15).

Existing NLP research has explored methods to augment these ambiguous claims with more contextual information. These include rewriting sentences to maintain the same meaning by summarizing previous text \citep{Choi2021-bm} or retrieving knowledge from an external database \citep{Wu2021-rt}. However, these techniques have not yet been implemented in current fact-checking tools. In this study, we discuss design elements necessary for the practical implementation of claim disambiguation. Participants suggested NLP tooling here might: 1) identify the spans in a claim that need disambiguation, 2) formulate queries for additional information required, 3) retrieve the relevant information, and 4) rephrase the claim accordingly. Complex or ambiguous arguments could also be decomposed into simpler sub-claims using question generation models with human supervision \citep{Chen2022-yz}. Inference beyond the sentence boundary may require anaphora resolution\footnote{Anaphora resolution requires resolving references to entities through different forms in a sentence (i.e., pronouns, abbreviations)} \cite{poesio2023computational} and entity-linking\footnote{Entity linking helps to identify the unique entities (i.e., people, places, government bodies) in text} methods \cite{petroni-etal-2021-kilt}. While the recent advances in NLP show promise for these tasks, their effectiveness in fact-checking requires further investigation. One approach is to formulate an effective workflow for Human-AI collaboration. For example, queries might be handled internally to the tool, externally via another tool (e.g., search engine, knowledge base), or provided and rephrased via a human-in-the-loop (HITL) approach \citep{demartini2020human}. A dataset consisting of original and rephrased claims would also be invaluable to train and evaluate tools for this task \citep{Choi2021-bm}. 

Disambiguation tools might falsely rephrase claims. To avoid this issue, P1 envisioned that a disambiguation tool must provide references used for generating the new claim, which makes it easier for fact-checkers to conduct sanity checks. Additionally, P5 said that even if the tool cannot successfully rephrase a claim, it would still be useful to at least highlight the parts of the claim that need further clarification.

\subsection{\textbf{Personalized Claim Filtering and Selection}}\label{idea5} 

Because the number of claims that could be checked far exceeds the capacity of human fact-checking, identifying the most checkworthy claims is crucial \citep{Sehat2023-xa}. Moreover, the diverse criteria used to gauge checkworthiness, and their relative emphasis, varies across individual fact-checkers as well as their organizations\added{ (Details of such differences are described in Appendix \ref{appendix:criteria-3})}. Examples of such criteria include: severity, topic, location, political leaning, targeted audience, etc.  \citep{graves2018boundaries}. While prior NLP work has formulated a checkworthiness prediction task \citep{alam2020fighting, Barron-Cedeno2023-py}, the one-size-fits-all nature of this formulation neglects the variance described above.

Consequently, participants expressed a desire for greater control and personalization in claim filtering and selection tools (See Appendices \ref{appendix:criteria-1} and \ref{appendix:criteria-2} for a full list of personalization criteria). For example, some participants (P2-3, 8-9) believed that a claim should be more checkable if there is already publicly available evidence, and thus should be ranked higher by a claim monitoring tool. This would require information retrieval of related evidence in scoring checkworthiness.
Similarly, P1 and P21 wished to compare the {\em spread velocity} across different claims. If a claim reaches a certain group of audience faster than other claims, P1 and P21 may be more interested in checking that claim. This would require creating and integrating a model of such claim spreads. 

Participants also suggested labeling claims for checkworthiness based on individual or organizational interest.  Labels would be collected through fact-checker daily work to train personalized prediction models. They also discussed providing manual feedback to prediction models in order to similarly personalize them. P12 said she wanted to be able to weight different criteria and let the model learn from her feedback. Similarly, P8 expressed a willingness to ``remove or adjust the checkworthiness ranking if a claim is not worthy of a high score.'' However, P8 also expressed concern about such a trained model suffering distribution shift over time: by labeling ``claims... made immediately after a breaking news event, the model might weigh related claims as more checkworthy'' long after that story was no longer newsworthy. 

Though expressed with regard to claim selection, a more general theme regarding human-AI interaction is personalized user training and calibration. For example, to make sense of the checkworthiness scores predicted by AI, P20 and P21 suggested an onboarding process. As P21 remarked, ``the combined [checkworthiness] score will be helpful once we have used the tool enough to set our own threshold for those ratings.'' Regarding human-AI calibration, P20 also mentioned that because ``there's some room for error that's taking into account these [AI] scores'', he would like the AI to convey how reliable its predictions are, such as by sharing its error rate akin to model cards \cite{mitchell2019model}.

\subsection{\textbf{Personal Bias Warning System for Claim Selection}}\label{idea6} 

Participants were highly conscientious regarding the risk of personal biases influencing their work (see Appendix \ref{appendix:criteria-1}). Repeatedly, there was mention of the importance of maintaining strict impartiality, despite acknowledging the loftiness of this ideal. For example, P1 remarked, "there were definitely examples where I think my own personal interests and backgrounds kind of led me to practice that over some more local stuff that would have been harder for people to individually fact check ... maybe that's [an area] where my own personal interests kind of affected my agenda.''

Similar to the media bias chart, which has been commonly used by journalists \citep{sheridan2021should}, participants envisioned an AI tool that could be an active partner in recognizing, classifying, and quantifying personal biases in claim selection. Key features of this tool included: 1) making fact-checkers aware of topical distribution of the claims they have selected (P4, 5), 2) flagging any potential bias towards selecting claims to fact-check depending on their source (e.g., political, regional, or racial) (P1), and 3) identifying claim selection patterns of individual fact-checkers by clustering them across attributes of interest (P1). 

P5 was positive toward such an AI tool, noting, ``we talk about the algorithm bias as well, [as] the way they [were] built, but I think the machine will be better than human in detail, or in not being as bias[ed].'' P4 also thought a tool that ``not limiting us in terms of topic area would be super helpful.'' Such a `bias detection' AI could help participants reflect on their own personal biases in claim selection, backing up assumptions with tangible data, and enabling them to counteract such biases. However, some participants also worried about how such a tool would produce results that might misrepresent the real situation if it takes a simplistic assumption about biases. For example, P15 said ``if all the false claims are coming from one party and not the other party, then we shouldn't try to create a false equivalence.''

Currently, NLP researchers use media bias charts\footnote{\url{https://www.allsides.com/media-bias}} and crowdsourced political ideology ratings to create datasets to determine the political ideology of articles and news outlets \citep{Chen2020-fg}. Models trained on these datasets could be used to categorize sources collected by a fact-checker and to examine whether these sources are neutrally selected. Topical analysis could also help find common themes from claims selected. Note that identifying and mitigating personal bias is challenging, such as oversimplifying the bias problem by forcefully trying to restore a balance of coverage, as P15 mentioned above. Further research is required at the intersection of bias and fact-checking more broadly. We summarize other aspects of potential biases mentioned by our participants in Appendix \ref{appendix:criteria-1}.

\subsection{\textbf{Human-AI Teaming for Fast Generation of Fact-check Briefs}}\label{idea7} 

Participants described fact-checking {\em briefs} as short, early drafts used internally to pitch and negotiate which claims would be fact-checked (P8, 9, 15). As P9 described it, a typical brief consists of~~``a rough indication [of] what the topic is, what we think the verdict is, and here's what we want to do.'' P14 begins fact-checking by first creating a fact-check brief, then pitching it to her editors in a team meeting before gaining approval to pursue a more detailed investigation. P15, a fact-checking intern, said that one of her main tasks is to prepare fact-checking briefs for senior fact-checkers. P9 and P15 both imagined that the AI can help improve efficiency by generating briefs in the same format as human-prepared briefs. While a prior NLP study has investigated automatic generation of fact-checking briefs \citep{Fan2020-sj}, the notion of briefs in that work (designed for crowd workers) is disconnected -- in content, format, and use -- from the notion of briefs being discussed here as used by professional fact-checking organizations.

Beyond AI-generated briefs, participants also envisioned a hybrid, human-AI collaborative approach. In this vein,  participants imagined that AI could generate questions to help fact-checkers research a topic from multiple angles (closer in spirit to \citep{Chen2022-yz} than \citep{Fan2020-sj}). For example, imagine one wished to fact-check the following claim: ``People in the Muslim Brotherhood openly stated they want to declare war on Israel'' (taken from \citep{graves2017anatomy}). P6 said that potential AI-generated questions that would ``essentially give [her] the next steps'' in her research process include: ``Who in the Muslim Brotherhood claim this statement?'' and ``When and where is the claim made?'' 

Beyond aiding negotiation of which fact-checks to perform, fact-checkers also envisioned a larger role for such briefs in helping to guide the ultimate fact-check itself. P15 wanted the AI to retrieve evidence from different sources, so she could then decide which directions to pursue. Participants also noted that questions lacking sufficient evidence could be a signal indicating the claim to be false. Additionally, P4 thought this AI could help in ``reducing the amount of time it takes sifting through information to find something that's relevant,'' as well as helping her ``think about things that we might sort of take for granted as known already'' to avoid self-presumptions (related in spirit to detecting and mitigating self-bias, as discussed in Section \ref{idea6}). 

\subsection{\textbf{Identifying Official Databases and Formulating Queries for Verifying Quantitative Claims}}\label{idea8}

% If a database can directly address quantitative claims, it is possible to develop an end-to-end claim verification system 
% \citep{Karagiannis2020ScrutinizerAM, Saeed2021-ff}. 
Access to statistical information is essential for fact-checking quantitative claims. Checking complex claims that merge statistics from various sources requires manual effort to contextualize these statistics by examining data from different sources. Participants noted that identifying the appropriate database is often a time-consuming task (P2, P18). P2 said, ``[it] slows us down ... it's a long drawn out process.'' To address this issue, participants imagined a tool that could help infer the official databases required to fact-check the claim and formulate queries for them (e.g., government databases and research statistics published by reputable agencies). For example, when a news article cited data from the Center for Disease Control and Prevention (CDC), P18 asked ``what [specific] database do I need to go to [from the CDC]? And what parameters do I need to input in order to get the data that was just cited by the New York Times?''

% \replaced{As a solution, participants}
Participants imagined an NLP model to first detect the specific parts of a text referring to quantitative data, and then generate queries to verify that data (including any additional metadata, such as location and date). For example, for the claim, ``the Supreme Court allowed warrant-less home searches within 100 miles of the US border'', outputs of interest would include: 1) labeled and extracted parts of the claim, 2) what original source the article refers to, 3) possible source of the statistics (i.e., public warrants issued by Supreme Court), and 4) other metadata (i.e., warrants effective date).

Building such an AI model would require integration across several standard NLP tasks. First, a sequence-tagging model \cite{bhattacharjee-etal-2020-bert} is required to detect quantitative parts of a claim and tag referred sources. Then, a generative model could be used to construct necessary queries for verifying the claim using that sequence tag from the tagged spans. To build and evaluate such a tool, a supporting dataset would provide a structured output containing: a) the official source, b) the exact name of the database to be consulted, and c) related metadata.

Recognizing the complexity involved, participants also acknowledge that a solution need not be fully automated to be useful. 
% not envision any obvious fully-automated solution during co-design. 
Specifically, P18 pointed out that using the AI tool to track and extract the sources of information is intended to simplify the search process, not to have AI directly provide answers to fact-checkers. They stressed that fact-checkers are still required to confirm the origins of all information themselves. During the discussion,
NLP researchers also suggested a HITL approach \cite{demartini2020human}, such as engaging online crowdworkers to help find desired information across diverse web databases. This would reduce work for professional fact-checkers while using crowdworkers to scaffold desired functionality until an automation solution becomes feasible.

\subsection{\textbf{Understanding Reader Engagement for Fact-checking Reports}}\label{idea9} 

Fact-checkers dedicated to curbing the spread of dis/misinformation care deeply about measuring the reach and impact of their work. Like other journalists, they want to understand who their readership is (and is not) and how they engage with posted content. Given the abundance of empirical studies that analyze reader engagement of fact-checking in the context of academic journalism \citep{Kim2022-ee, Robertson2020-fh}, there is a corresponding need for advanced tools to assist professional fact-checkers in analyzing reader behavioral data. Most participants reported using web analytics tools (e.g., Google Analytics\footnote{\url{https://developers.google.com/analytics}} and Trendolizer\footnote{\url{http://get.trendolizer.com/}}) to measure the reach and spread of their work, but these tools are often limited to examining the readership of fact-checking reports. 

Our participants (P4, 7-9, 13) wanted to analyze how readers of different demographics engage with their work, and to compare the impact of their work to other fact-checkers within or across organizations. Regarding the readership composition, P7 wanted to know,  ``what areas it's spreading or the common characteristics of people who are engaging with... [Theoretical targeting] really helps us reach different groups of people while spreading our accurate information so we try to change our narratives to reach [more] people ... [or] you can just maybe convince people [more] easily.'' P4 echoed that, ``We have a claim, and we want to have some storytelling about the contextual information surrounded, and what kinds of potential audience, demographics, attraction, and attention it attained. On social media, this figure is very important.'' Akin to search engine optimization (SEO), such understanding would enable fact-checkers to make specific changes to their fact-check reports to better reach and impact their intended audience(s). 

As noted above, measuring impacts on live traffic is common by instrumenting webpages to collect analytics quantifying user engagement with content. However, this only works after content has already been posted, and conducting A/B experimentation of variant presentations with live traffic is always risky. Conducting surveys or focus groups with target audiences provides more focused and qualitative feedback, but is also post-hoc and involves greater latency. 

Akin to forecasting disinformation (Section \ref{idea1}), forecasting spread and audience reactions (as a function of fact-check content or presentation) has potential to enable fact-checkers to predict such impacts before content is posted, and to tailor it to optimize impact. For example, network analysis can be performed to predict spread \cite{Zhang2021-mh}. Other work has correlated content analysis of text with audience reactions to guide best practices in general \citep{Kim2022-ee}. Perhaps most intriguing from an NLP perspective is recent work seeking to simulate the reactions of different demographic groups via large language models \cite{argyle2023out,kim2023ai,santurkar2023whose}. Such work is quite nascent but early work is encouraging.  

\subsection{\textbf{AI Assistance in Writing Fact-checking Reports}}\label{idea10} 

Participants imagined NLP tooling to aid them in writing fact-check reports, especially the repetitive, standardized portions. As P3 put it succinctly, ``We spend too much time writing and copying.'' With the rapid rise and spread of large language models (LLMs) such as GPT-4, it has become increasingly common to have humans revise AI-generated texts rather than writing texts from scratch. To do so effectively, researchers found that LLMs must be tailored to meet domain-specific needs (i.e., fact-checking) and a claim-driven extractive step helps improve abstractive summarization \citep{Russo2023-qd}.

Because fact-check reports usually have a consistent writing structure within organizations (see Appendix \ref{appendix:criteria-2}), fact-checkers first want to specify the structure, then use AI to generate content accordingly. For example, P2 wanted AI first to generate descriptions of who said the claim and the source, such as ``say it in a tweet that ...the White House claimed that...,'' then explain the evidence ''there was no vaccine available at the time...'' More generally, participants (P3, 9, 19, 20) stated they would like to provide the claim, its verdict, and the evidence they have collected to verify the claim. The generated report would contain a title, ClaimReview\footnote{\url{https://schema.org/ClaimReview}} tags, an archived URL of the claim, blockquote fields linking the claim, and captioned images with the verdicts. Additional desired features included producing a description of that claim, possible suggestions for writing supports or counterarguments to the claim, and synthesis of any other raw information provided by the checker.

While today's LLMs can take natural language prompts as input and provide suggestions for writing different parts of a report, open research challenges include performing complex and implicit multi-step reasoning (e.g., generating fact-checking reports \cite{Chen2023-fu}, retrieving relevant and accurate external information to enhance generation \cite{lewis2020retrieval}, appropriately citing such evidence \cite{Kamalloo2023-nz}, and enforcing other specific formatting requirements).

\subsection{\textbf{AI Assistance in Critiquing and Editing Fact-checking Reports}}\label{idea11} 

Whereas the previous idea envisioned AI generating a rough copy of content for humans to refine, this idea reverses these roles, with humans writing polished content and the AI taking on the role of the human editor, critiquing and editing content. Inspired by our 3rd AI probe to review the argumentation quality of a fact-check, participants envisioned that AI models could help with judging their writing quality. They imagined that a model could reliably judge fact-checking reports across dimensions such as readability, coherency, clarity, and quality of the argument presented (more such criteria is listed in Appendix \ref{appendix:criteria-2}). P4 said, ``if we’re writing in English, everything has to be accessible to people who have a limited educational background and whose first language isn’t English.'' P6 also echoed that ``[you need to] ensure that concepts are thoroughly explained. Inflation, for example, not just talking about it [as a concept], but actually explaining what, when, and how.''

Moreover, the participants imagined AI tools will take domain-specific content and presentation requirements into consideration and allow customization to capture their organization's writing constraints. For example, P9 highlighted that: ``every organization seems to present their checks in different styles. If there was some flexibility in this model, where you could input your own organization's style that would be cool.''

Participants were also interested in subjective aspects of the writing, such as, the strength of presented evidence or pointing out silos in their arguments. P9 said, ``[the tool can check] if all aspects of the claim have been supported with clear evidence and they are clearly linked together.'' Moreover, P8 said because journalists can sometimes be oblivious to their confirmation biases, if AI could provide adversarial challenges to the written arguments, it could potentially improve the reporting to maintain greater objectivity.

% Participants also mentioned more subjective aspects of writing fact-checking reports, 

While the use of LLMs to generate draft content is increasingly popular, the idea of AI automatically critiquing and refining professionally produced, polished content appears to be more novel, challenging, and subjective. In contrast with familiar automated checking of spelling and grammar, and perhaps even automated student essay grading in standardized testing, sophisticated editing of professional content would seem to take such AI-assistance to an entirely new level. Further consideration of domain-specific content and presentation requirements would make this task even more challenging yet.

Current research in AI-assisted writing has looked into more subjective domains such as story writing \cite{Lee2022-mk, Wu2022-pd} by using LLMs. Prior research \cite{Lee2022-mk, Wu2022-pd} suggests the promise of using LLMs as effective collaborators for different writing tasks. However, more research is required for AI-based critiques for subjective aspects of a fact-check report. As a starting point for editing fact-checks, one could explore few-shot learning \cite{brown2020language} with a relatively small expert annotated parallel data (a corpus that contains both the initial text and the edited version of the text) that helps edit the initial text to a more-polished version appropriate for fact-checks.

\section{Discussion} \label{sec:discussion}
In this section, we discuss the benefits, efficacy, limitations and improvements of \textit{Matchmaking for AI}, including its implications for human-centered fact-checking and general AI co-design. First, we present a summary of co-designed ideas for fact-checking to illustrate their novelty, extending previous empirical studies. We also explain that \textit{Matchmaking for AI} is an example of translational HCI practice as we produce design ideas useful for practitioners. Additionally, our findings suggest that matchmaking helps brainstorm ideas with concrete design suggestions and relevant technical details. Finally, we articulate its limitations and future improvements, helping researchers better adapt this co-design approach to other AI application domains.    

\subsection{Benefits of Co-designing AI via Matchmaking}
\label{subsec:benefits-matchmaking}

\subsubsection{Summary of Co-Designed Ideas for Fact-checking}\label{subsec:sumamry-of-ideas}

In our study, fact-checkers generated new design ideas that address various needs that arise across the fact-checking process. These ideas either satisfy novel needs that were not mentioned in prior ethnographic studies or offer design solutions for existing needs documented in prior work, which existing tools or NLP research do not address. 

The majority of co-designed ideas assisted in information searching, processing, and writing tasks for efficient and personalized fact-checking. Much of existing NLP research focuses on supporting claim selection and automatic claim verification. In our study, fact-checkers also generated an idea related to claim selection (Idea 5), highlighting important yet missing selection criteria that these tools must accommodate. Other co-designed ideas covered tasks in the fact-checking workflow beyond claim selection. Fact-checkers wanted to leverage NLP to identify broader disinformation narratives (Idea 2), to dynamically monitor the credibility of social media users (Idea 3), to find and provide context for ambiguous claims (Idea 4) so that fact-checkers can process information more efficiently and make a judgment on the credibility of the claim themselves. Writing the fact-check briefs and reports was another important area where fact-checkers thought NLP (and human-AI teaming) could make their work efficient (Ideas 8-11).

Our co-design session also revealed needs to support diverse goals of fact-checkers beyond conducting the fact-checking task. They wanted help in forecasting disinformation (Idea 1) so that they could proactively prepare resources to combat future misinformation. They also wanted assistance in monitoring their own potential biases in claim selection (Idea 6). Finally, because much of fact-checking work is now done by a group of fact-checkers, they thought that NLP could help them quickly generate fact-check briefs for internal discussion on what misinformation to prioritize and focus on (Idea 7). \added{Additionally, we learned that fact-checkers’ needs and perceived importance of the co-design ideas might differ depending on their regional focus and organizational interests. We describe these exploratory findings in Appendix \ref{appendix:criteria-3}.}

\subsubsection{Matchmaking Addressing the HCI Translational Gap}

HCI scholars have discussed the importance of producing translational work to bridge the gap between research and practice \citep{Colusso2019-ti, Norman2010-fh, Churchill2020-mg, Colusso2017-lh}. Our study is an example of translational HCI practice for fact-checking. Matchmaking for AI translates knowledge from previous empirical fact-checking research \citep{graves2017anatomy, micallef2022true, juneja2022human} to establish AI design requirements, and imagines useful fact-checking tools by adopting state-of-the-art AI techniques. By doing so, we address the misalignment between the practices of AI-based fact-checking driven by technological challenges and fact-checker needs, practices, and values. Our results encourage building, human-centered AI tools that are helpful to fact-checkers. 

As noted by both \citet{Colusso2019-ti} and \citet{Churchill2020-mg}, ethnographic research in HCI produces theories and thematic understandings to explain human practices and problems of using technologies. This knowledge often fails to describe needs with sufficient detail required for design. Additionally, practitioners rarely have time to read ethnographic studies and are often unfamiliar with academic theories, concepts, and terminology. \citet{Norman2010-fh} concludes that practitioners want to understand: 1) what stakeholders actually need; and 2) how the existing tools can be improved for their intended use, or adapted to serve different use cases which they were originally intended for. 

Matchmaking for AI helps stakeholders formulate ``need-based statements'' (written at the beginning of each idea) to be useful for practitioners. In Step (1) (see Section \ref{subsec:domain_expertise}), participants are empowered to articulate what needs to be done to meet their desired outcomes and criteria specialized in each separated fact-checking task. Thus, these “need-based statements” have clear goals, reducing potential design conflicts for practitioners. Additionally, because we conduct Step (1) as a mapping activity by guiding participants in small steps to “construct and express deeper levels of knowledge as their experiences” \citep{Visser2005-nt}, participants help uncover new needs (summarized in Section \ref{subsec:sumamry-of-ideas}) that are not directly identified from the prior work.

In Step (3) (see Section \ref{subsec:codesign_feasible_AI}), participants prioritize the most important needs to be addressed and co-design ideas with NLP researchers to address these needs. First, they brainstorm add-on functionalities to improve existing tools for their intended use. For example, in Idea 5 (Section \ref{idea5}), participants brainstorm concrete functions that enable them to flexibly filter and select claims on existing claim selection tools. Also, they update general annotation schemes \citep{Konstantinovskiy2021-od} of claim detection based on organizational interest. Additionally, they point out new use cases that both designers and NLP researchers have not thought of, inspiring new adoption of AI techniques, such as using fake news generators to forecast disinformation ahead of an event (Idea 1, Section \ref{idea1}) and using common narrative patterns as queries to retrieve previously checked or group unchecked claims (Idea 2, Section \ref{idea2}). 

In summary, our work complements \citet{Colusso2019-ti}'s translational model by employing Matchmaking for AI as an example of translational practice in fact-checking. We produce design ideas with different levels of maturity, providing a ``north star'' to guide practitioners for practical adoption.

\subsubsection{Implications of Co-designed Ideas on NLP Research}

Interdisciplinary research across HCI and NLP has shown the importance of incorporating human feedback into AI development and evaluation \citep{Wang2021-cv, Lertvittayakumjorn2021-rh}. Previous studies \citep{Lai2021-vg, Das2023-tq, Subramonyam2022-am} also describe that empirical evaluations of human-AI decision-making require more fine-grained metrics to measure AI performance (e.g., accuracy, efficiency, and calibrated trust). These metrics should be framed by stakeholder domain expertise so mixed-initiative AI systems can be more reliably measured \citep{Cai2019-xb}. In this study, we provide specific and nuanced definitions of these metrics, which could either improve model development or evaluation. For example, if AI researchers want to build models to forecast social media engagements for online misinformation, they need to carefully consider what it means for a social media post to be \textit{viral}. How can we, as researchers and developers, quantify a \textit{viral} post and allow customization for different contexts? In Appendix \ref{appendix:criteria-1}, we describe four unique definitions of virality mentioned by different participants, including virality between social media vs. traditional media, attention raised by public figures, and sensationalized or hyperbolic language that grabs public attention or is easily misinterpreted. These details of often abstract and ill-defined values help developers refine the annotation scheme for labelling viral claims across different media environments and operationalize metrics to test AI performance between in-distribution and out-of-distribution datasets.

\subsection{Efficacy of Co-designing AI via Matchmaking} \label{subsec:efficacy-matchmaking}

\subsubsection{Fostering Effective Communications with NLP Experts for Idea Brainstorming} \label{subsubsec:foster-feasible-ideas}
While prior co-design produces high-level design guidelines or imaginary AI solutions, their outcomes often lack concrete design suggestions and relevant technical details. By bringing NLP researchers into stakeholder conversations, matchmaking constructs a progressive idea brainstorming flow for participants to: 1) explore the maximum capabilities of AI; 2) align AI capabilities and limitations with human criteria; and 3) propose applicable design recommendations.

First, participants brainstormed new AI possibilities with a better understanding of what existing AI offers. For example, some participants who had used off-the-shelf claim detection tools noticed high AI accuracy in determining whether a sentence constitutes a claim. NLP researchers explained that because a claim syntax can be well defined with objective annotations, a claim detection model gains a high accuracy by training on a gold dataset. Thus, participants (P3, 8, and 14) imagined NLP tools might further identify claims that are politically aggressive, provoking emotions like fear or calls to action. They presented exemplar claims containing these attributes and discussed with NLP researchers how to annotate these semantic accurately (results in Appendix \ref{appendix:criteria-1}). Additionally, while experimenting with the AI Probe reviewing the argumentation quality of fact-checks, P9 compared it with the existing commercial writing assistants. He proposed designing a similar AI assistant to generate a meta-review of the fact-checking percentage that successfully addressed different ambiguous aspects of a claim and how persuasive the fact-check explains evidence (see Idea 11, Section \ref{idea11}).

Meanwhile, by knowing where and why AI might possibly fail, participants align AI capabilities and limitations with human criteria. For example, many participants (P8, 20, and 21) discussed with NLP researchers how to align AI decision thresholds and errors of claim check-worthiness with humans. As previously mentioned in Idea 5 (Section \ref{idea5}), identifying a single, most check-worthy claim from a large volume of claims requires detailed examinations on different selection criteria. However, participants reported that results from existing claim selection tools often fail to meet their standard. This dissatisfaction might be attributed to the issues of trust and the lack of AI explainability. Because selecting which claims to check first is a critical task for fact-checkers, AI decision-making process should be explained so that they can discern how and why certain claims were prioritized over others \citep{Das2023-tq}. This process helps promote appropriate trust by aligning AI predictions with user expectations \citep{Lee2004-kt, Ma2023-rj}.

During the workshop, participants suggested several design recommendations, including the use of AI explanations, to foster human-AI appropriate trust. For example, while iteratively testing different inputs of the AI Probe detecting social media check-worthy claims, P20 proposed to set a human-AI decision threshold by aligning the model decision and its error rate with human judgement. Only when the AI predicted score is over a minimal human-AI threshold would he look at AI-suggested claims. Similar to P20, P21 said once fact-checkers have used the tool enough to set their own threshold, such as the highest threshold for worthiness and unworthiness, the AI advice would become more effective and useful for humans. Additionally, participants (P8, 14) brainstormed interactive functions to adopt feature-based AI explanations for Idea 5 (Section \ref{idea5}). After learning that certain parts of the claim exert a greater influence in the final AI prediction, P8 proposed to rank claims and highlight important parts of the claim to explain its ranking: "I think ranking them is even better, then you could highlight important texts explaining why something has been given a certain priority over other claims. Potential harm [is a priority] because it's [highlighted as] a health claim." 

\subsubsection{Implications of Stakeholder Algorithm Aversion on AI Co-design}
During our matchmaking, many fact-checkers initially exhibited algorithm aversion \cite{Dietvorst2015-eq, lee2018understanding}. We found that having participants experiment with AI Probes and communicate with NLP researchers helped reduce their aversion and restore a constructive flow of idea brainstorming.

Observed levels of algorithm aversion varied significantly across participants. For example, many participants (P2, 4, 6, 11, 12, 15, 16) believed that algorithm accuracy is inferior to human accuracy, so human oversight would be imperative when using AI. While talking about the AI Probe that robo-checks quantitative claims, P12 wanted to further verify those statistics himself: ``I just want to make sure it wasn't pulling numbers from inaccurate parts of whatever the document is''. P13 also highlighted that AI checking the writing quality of fact-checks is not comparable to editors’ eyes: "even if it looks promising, you [still] need a human eye in the form of a human sub-editor.'' Some participants (P5, 11, 16) were more hesitant to use AI because they were aware of certain AI biases. For example, P11 said “machine learning is the process of continuous learning by the computer itself, so what we put into the process [datasets] results in a certain bias.” Additionally, some participants (P5, 8, 9) expressed stronger concerns about over-relying on AI. P5 clarified, “the machine can support you, but you should not completely rely on it.'' 

Although participants exhibited some algorithm aversion, we observed a changing perspective from matchmaking Step (1) to Step (2). For example, P5 who was previously averse, expressed his willingness to use AI after experimenting with the AI Probes. In Step (1), he said, "Humans can be biased, and I'm aware [of that], but I think what's more [severe] is algorithm bias." After P5 saw example outputs from the AI Probes, he thought AI could perform certain tasks accurately, such as capturing claim syntax (e.g., classifying if the text is a claim, an opinion, or a prediction) and identifying quantitative information. Similar to P5, P16 was previously hesitant to have AI monitoring claims because he thought AI is “rigid” and only humans can flexibly determine what is check-worthiness based on the local news context. But he was interested in the AI Probe checking the argumentation quality of a fact-check: ``I think this can be trustworthy. But when using this tool, we still need editors to touch on these copies.''

As more HCI research explores AI co-design, we believe that participant aversion to AI could be either a barrier or an elixir, influencing the quality of co-design outcomes. Although this aversion might lower participant initial interest, framing it as a design problem can, in turn, elicit participation. As suggested by \citet{Hou2021-kp}, the transition from algorithm aversion to appreciation requires researchers to provide {\em Expert Power} for users (i.e., the result of being more knowledgeable, competent, and knowing what better action to take). As shown in our ideas and participant design experiences, Matchmaking for AI helps participants cultivate such Expert Power by engaging them in active conversations with AI experts. This reflects the traditional human-centered design principle held by the design community that user trust in technology should be mutually communicated and agreed on between users (who are served by the technology) and developers (who create the technology), rather than singly produced or transferred from either side \citep{Steen2013-ut, sanders2008co, Norman2010-fh}.

\subsubsection{Implications of Stakeholder-centered AI Design in Fact-checking}

Recently, there is an increasing trend in HCI and design research to use stakeholder-centered AI approaches. The goal is to more effectively tailor AI development to meet the specific needs of various domain-specific stakeholders. In our study, we focus on a specific group of stakeholders -- fact-checkers. Due to their varied backgrounds, the ideas we co-designed with them display a range of AI literacy and acceptance levels. We acknowledge that this diversity influenced by their cultural perceptions and sensitivities towards AI might lead to uneven adoption rates for these ideas in the future. To enhance this, we suggest future stakeholder-centered AI design for fact-checking could be developed in two directions: 1) employing an open-sourced design strategy to build AI infrastructure and practice; and 2) using an iterative design process to customize AI solutions, taking into account their unique cultural contexts. We describe these two directions as follows.

Compared with other stakeholder-centered AI studies that focus on designing AI tools for gig-workers (e.g., Uber drivers \citep{zhang2022algorithmic}, food courier \citep{Ma2023-dm}), public services providers (e.g., frontline workers \citep{Kuo2023-ms}), or other domain-specific specialists (e.g, clinicians \citep{Kim2024-hp}), we point out that there is a unique opportunity for fact-checkers to develop common AI infrastructure around the globe based on their international network and collaborative practice. In our study, participants frequently mentioned that the existing International Fact-checking Network (IFCN) helps establish a code of conduct for a standard fact-checking workflow. Additionally, some of the existing computational infrastructure was also grounded by this collaborative work, such as the CoronaVirus check database\footnote{\url{https://www.poynter.org/ifcn-covid-19-misinformation/}} and the ClaimReview tagging system\footnote{\url{https://www.claimreviewproject.com/}}. Thus, co-design ideas generated from our study could be further developed as similar open-source AI infrastructure.

In our study, \textit{Matchmaking for AI} is employed through a linear design process to gather fact-checker initial AI needs and ideas. However, design in practice is often non-linear, involving iterative reframing of both the design goals and the solutions through a deeper comprehension of practical problems \citep{Zimmerman2022-xy}. We argue that, bearing this design nature in mind, our design ideas could spur diverse AI products by tailoring them into fact-checker unique cultural contexts.

\subsection{Limitations and Improvements for Future Matchmaking for AI} \label{subsec:limitations}

% As a reflection, we found that the existing process of 
In this section, we discuss limitations of \textit{Matchmaking for AI}.
% presents limitations that should be improved in future adaptations. 
For example, our protocol requires considerable time. We designed a three-step process, aiming to understand participants work practices, have them understand AI, and brainstorm ideas based on this understanding. However, during the final brainstorming session, we had time to develop around two ideas only. We may not have captured all useful ideas due to this time constraint.
% may have led to This means that we might miss other useful ideas. 
We posit that if co-design participants have more cross-domain knowledge, more time could be allocated to the brainstorming session to design more workable ideas.
% Future studies should further adjust time allocation tailored to stakeholder and researcher knowledge about the domain. 
% For example, if both fact-checkers and AI researchers know each other the domain, Step (1) and (2) could be shortened.} 
% Similarly, if participants possess AI knowledge, Step (2) could focus on directly discussing state-of-the-art AI techniques without providing AI Probes.}
% Design goals should be sufficiently clarified for the matchmaking Step (3). As a reflection, we did not articulate the design goals clearly to participants -- whether exploring a new design space for NLP that prior work doesn't address or any design ideas that address user needs regardless of the novelty. This results in some ideas that participants brainstormed were not very novel, which we chose not to include in the table \ref{tab:nlp-based fact-checking solutions} and put them in the appendix \ref{appendix:existing-needs-designs}. Thus, we recommend that facilitators clearly defining the criteria for ideation could better guide the brainstorming process and also enable effective time management.}

In our matchmaking step, we could have communicated upfront that the goal of the brainstorming session was to design novel ideas that were both attuned to their needs and not already addressed in prior NLP literature. This indicates that co-designers should have helped participants understand what was and wasn't novel. Not doing so may have led to some design ideas that were less novel (see ideas in Appendix \ref{appendix:existing-needs-designs}). Similarly, we may have allowed too much co-design time to be spent discussing ideas that were not novel.

% Moreover, in the preceding paragraph we discussed a dearth of time during the brainstorming session which may have been a by product of this gap.}
% Thus, we recommend future researchers clearly defining the criteria for ideation.

% In our matchmaking step, we did not communicate the goal of the brainstorming session with the participants upfront, i.e., to design novel ideas that attuned to their needs but were not addressed in prior NLP literature. 

There was also a missed opportunity for participants to review and prioritize these ideas as part of the matchmaking process, since participants were not interviewed again afterward. Moreover, some ideas received more time and attention than others, especially when multiple participants mentioned them. This led to an uneven level of design details for the ideas. Future work could take a more structured approach to alleviate this issue by a) gathering as many ideas as possible in the beginning, b) presenting all ideas to participants for importance rating, and then c) conducting the existing matchmaking Step (3).

% Moreover, we met the participants individually to brainstorm different ideas
% Since we met the participants individually to brainstorm different ideas, 
% some of the ideas might have more details than the others when multiple participants mentioned the same idea idea. To improve our protocol in the future, a more structured approach could help by a) gathering as many ideas as possible, b) presenting all ideas to participants for importance rating, and c) conducting the existing matchmaking Step (3).}  

Running remote co-design workshops enabled us to engage with fact-checkers from around the world, and to more easily recruit busy professionals with limited availability. However, unlike traditional workshops wherein a group of participants brainstorm ideas together through sketching, mapping, or other design activities, the remote setting made it more challenging for participants to engage interpersonally and express their creativity. In future work, it would thus be interesting to explore \textit{Matchmaking for AI} in group and/or in-person settings, which might also benefit researchers in analyzing participant ideas with a regional focus.

While we successfully developed a set of workable NLP-based ideas, our explicit focus on NLP reflects our own research bias. Due to this, we knowingly steered participants more toward ideas related to NLP-based approaches. This may have led, for example, to omitting ideas related to multi-modal content, which is important to today's fact-checking landscape.
% Throughout the study, the facilitator might not encourage participants as much to brainstorm ideas related to multi-modal content, omitting novel ideas in this field. Undeniably, checking misinformed multi-modal content is crucial in today's fact-checking landscape.
In addition, we have employed proactive strategies to closely align the study with journalistic values. However, we recognize that despite our best efforts, there may be gaps between our research and practical challenges in applying AI to journalism. We observed the presence of algorithmic aversion in participants, especially around the ethical concerns of incorporating AI solutions into practice (described in Section \ref{subsec:efficacy-matchmaking}). For example, participants were averse to using AI solutions for tasks that might heavily involve ethical and moral judgments. We emphasize that there is a need for an ongoing dialogue between AI researchers and journalism professionals to address the complexities of applying AI in fact-checking while maintaining human agency.
% or only allowing them to facilitate human mundane work. 
% as described in Section \ref{subsec: positionality}. 
% However, we recognize that these alignments might not fully bridge the gap between AI research and the practical realities of journalism,

% as some participants saying that a highly integration of AI into their workflows might fail to   the algorithmic aversion observed from participants (described in Section \ref{subsec:efficacy-matchmaking}) and reported in other research about automated journalism \citep{Lermann_Henestrosa2023-le}. We want to emphasize the need for ongoing dialogue and collaboration between AI researchers and journalism professionals to address the complexities of applying AI in fact-checking while maintaining human agency.}

Inspired by recent advances in LLMs, we envision that many future AI Probes could be implemented via LLMs. During our co-design studies, we improvised using GPT-3.5 as an additional probe for one of our final participants (P21). LLMs support a variety of useful NLP tasks with minimal burden for modeling and data acquisition \citep{Beltagy2022-fc, brown2020language}.  During our co-design workshops, P21 and facilitators used GPT to generate a misinformation template (Idea 1, Section \ref{idea1}) by specifying a news event in a ``zero-shot'' setup \citep{Beltagy2022-fc}. Experimenting with GPT was a starting point for this participant to describe more concrete design elements of a feasible AI solution without knowing extensive technical details. This suggests that LLMs could naturally serve as AI Probes that help non-AI domain experts envision new AI applications in their own contexts. For this reason, we believe that adopting LLMs is a promising direction to co-develop feasible AI solutions in future co-design research.

\section{Conclusion} \label{sec:conclusion}

To better understand the needs of professional fact-checkers and opportunities for AI-based assistive tooling, we investigated a novel form of co-design that brought together a tripartite group of fact-checkers, designers, and AI researchers. In particular, we proposed \textit{Matchmaking for AI}, extending \citet{Bly1999-ar}'s earlier {\em Matchmaking} concept to an AI co-design process. This co-design method seeks to accelerate the effective translation of state-of-the-art research into practice, to better support and benefit stakeholder or practitioners lacking AI expertise. Also, co-designed ideas from our workshops exemplify how fact-checker needs can or should be addressed by AI technology, and we uncover new needs not identified in prior empirical studies. Meanwhile, co-designed ideas offer a ``north star'' to guide future NLP-based fact-checking research toward technology development with greater potential for adoption by professional fact-checkers. Additionally, by providing detailed descriptions of the matchmaking process, as well as our reflections on its efficacy, limitations and future improvements, we illustrate the broader potential of our co-design approach for other domains beyond fact-checking. 

\section{Acknowledgements}
We appreciate the valuable writing feedback provided by the reviewers and the research insights from our external collaborators, including Tamar Wilner, Scott Hale, and David Corney. We also thank our participants, without whom our research would not be possible. This research was partially supported by the following: the Knight Foundation; the National Science Foundation IIS-1939606, DGE-2125858 grants; Good Systems\footnote{https://goodsystems.utexas.edu}, a UT Austin Grand Challenge for developing responsible AI technologies; and UT Austin's School of Information. The statements made herein are solely the opinions of the authors and do not reflect the views of the sponsoring agencies.

\bibliographystyle{ACM-Reference-Format}
\bibliography{reference}

% \pagebreak
%TC:ignore

\appendix
\section{Fact-checking Criteria - Part 1} \label{appendix:criteria-1}

\begin{table}[H]
\resizebox{0.88\textwidth}{!}{%
\begin{tabular}{p{0.2\linewidth}p{0.7\linewidth}}
\toprule
Criteria &
  Description (which participants voiced this criteria) \\ \midrule
\multicolumn{2}{l}{\textit{Monitoring and selecting claims}} \\ \midrule
News judgment &
  \begin{tabular}[c]{@{}p{\linewidth}@{}} What news sources are good for finding claims to check (3, 7, 10, 18)
  % \\ Plan topical news before head (2, 20)
  \end{tabular}
   \\ \midrule
Newsworthiness &
  \begin{tabular}[c]{@{}p{\linewidth}@{}}Breaking news that the public cares about (3, 5, 8)\\ Trending social media posts (10, 11)\\ Reported news by politicians or traditional media (3, 5, 11)\end{tabular}
   \\ \midrule
Fact-checkable &
  \begin{tabular}[c]{@{}p{\linewidth}@{}}Whether it is a claim, not an opinion or prediction (All)\\ Whether there's any publicly available evidence to validate (2, 3, 8, 9)\\ If the claim or claim's terms are ambiguous, they should be well-defined (3, 4, 9, 10)\end{tabular}
   \\ \midrule
Predicted truth &
  \begin{tabular}[c]{@{}p{\linewidth}@{}}Quickly evaluate whether the claim is true (2, 3, 9, 17, 19, 20)\\ Prioritize claims that are most likely to be false (7, 9, 16, 17, 20)\\ Primarily check false claims because organizations get paid from social media companies (9, 19)\\ Also check claims that are likely to be true (18)\end{tabular}
   \\ \midrule
Virality &
  \begin{tabular}[c]{@{}p{\linewidth}@{}}Trending claims on different social media platforms or reach threshold of engagements (1, 2, 5, 21)\\ Claims reported by traditional media outlets (13)\\ Claims repeated by public figures (8, 10, 13)\\ Claims that are sensationalized, or misinterpreted, or contain hyperbolic language, grabbing public attention (1, 3, 8, 14)\end{tabular}
   \\ \midrule
Harmfulness &
  \begin{tabular}[c]{@{}p{\linewidth}@{}}Claims specifically relating to medical issues (5, 6, 15, 19)\\ Claims containing risks on finance, public welfare, people's understanding of society (19)\\ Claims containing persuasive language that calls people to action (3)\\ Claims targeting a specific region or group of people (5, 9, 13, 15, 16)\\ Claims that are both viral and relevant to people's physical and social lives (1, 2, 4, 8, 11, 12, 19)\end{tabular}
   \\ \midrule
Personal Interest &
  \begin{tabular}[c]{@{}p{\linewidth}@{}}Claims that are less viral but specifically target certain groups of people (5, 9, 11)\\ Claim topics that fact-checkers are very familiar with (4, 6, 12, 13)\end{tabular}
   \\ \midrule
Organizational Interest &
  \begin{tabular}[c]{@{}p{\linewidth}@{}}Politics (2, 3, 12)\\ Region-specific (5, 9, 13, 15, 16)\\ Global (20, 21)\\ Topics previously with ample misinformation (4, 6, 8, 13)\\ Claims relating to young audience (18)\end{tabular}
   \\ \midrule
Biases awareness &
  \begin{tabular}[c]{@{}p{\linewidth}@{}}Intentionally try to find claims from different parties (5, 6, 8, 16)\\ Maintaining neutrality is essentially impossible (3, 15, 20)\\ Do not conduct themselves in any political affairs (3, 7, 19)\\ Editorial review for political biases (5, 7)\\ LGBTQ (11)\\ Intentionally hire diverse groups of fact-checkers (3, 18)\end{tabular}
   \\ \bottomrule
\end{tabular}%
}
\vspace{0.2cm}
\caption{Fact-checking Criteria - Part 1}
\end{table}
% \pagebreak
\section{Fact-checking Criteria - Part 2} \label{appendix:criteria-2}
\begin{table}[H]
\resizebox{0.88\textwidth}{!}{%
\begin{tabular}{p{0.2\linewidth}p{0.7\linewidth}}
\toprule
Criteria &
  Description (which participants voiced this criteria) \\ \midrule
\multicolumn{2}{l}{\textit{Investigating claims}} \\ \midrule
Provenance &
  Identifying claims origins to verify claims (2, 3, 8, 13)
   \\ \midrule
Reliability &
  \begin{tabular}[c]{@{}p{\linewidth}@{}}More than one source for validation (15)\\ Addition of multiple experts (10, 12, 13, 15)\\ Checking news source bias (8, 13)\\ Don't use secondary sources (5, 13)\\ Sources from reputable international, non-profit, or non-partisan organizations (2, 4, 5, 13)\\ Follow evidence pyramid for scientific facts (4, 6)\end{tabular}
   \\ \midrule
Thoroughness &
  \begin{tabular}[c]{@{}p{\linewidth}@{}}Ask contextual questions to breakdown claims by internal fallacy (2, 5, 13) \\ Define the exact wording of every term/concept (1, 2, 6, 8)\\ Reconstruct a claim's origin (3, 6, 7, 13)\end{tabular}
   \\ \midrule
\multicolumn{2}{l}{\textit{Writing fact checks}} \\ \midrule
Writing structure &
  \begin{tabular}[c]{@{}p{\linewidth}@{}}Begins with facts, introduces claim \& context, then evidence explaining the rating (2)\\ Begins with the claim \& context, then gives the verdict, \\ provides facts explaining that verdict, concludes with a final verdict (3, 4, 6, 8, 9, 13, 15)\end{tabular}
   \\ \midrule
Readability &
  \begin{tabular}[c]{@{}p{\linewidth}@{}}Simple language for readers to understand (4, 6, 8, 10, 11, 13)\\ Definition of a concept should be well-explained (2, 8, 9)\end{tabular}
   \\ \midrule
Coherence &
  \begin{tabular}[c]{@{}p{\linewidth}@{}}Clearly justify the provenance of the claim (3, 13)\\ Readers have access to all reliable sources to replicate the check (1, 2, 8, 10, 12, 18)\\ Enough evidence to validate the claim clearly (2, 4, 8, 9)\end{tabular}
   \\ \midrule
\begin{tabular}[c]{@{}p{\linewidth}@{}}Collaborative or \\ hierarchical editing\end{tabular} &
  \begin{tabular}[c]{@{}p{\linewidth}@{}}Peer review (5, 10, 11)    \\ 2 rounds of editorial review (2, 4, 6, 8, 9, 13, 15)\\ 3 or more rounds of editorial review (19, 22)\end{tabular}
   \\ \bottomrule
\end{tabular}%
}
\vspace{0.2cm}
\caption{Fact-checking Criteria - Part 2}
\end{table}
% \pagebreak

% \pagebreak

\pagebreak
\section{Existing Needs and Designs} \label{appendix:existing-needs-designs}

While our co-design process was primarily focused on generating novel ideas (Section \ref{sec:results}), four additional ideas were generated that remain significant and merit further attention, despite earlier coverage in prior work \citep{Procter2023-rr, micallef2022true, graves2018boundaries}. For these ideas, we synthesize participant quotes and our reflections. We expect this material to further inform community understanding of this need, and to further encourage additional technical and design work to develop, implement, and deploy these ideas in practice. 

% Please add the following required packages to your document preamble:
% \usepackage{booktabs}
% \usepackage{multirow}
% \usepackage{bbding}
\begin{table}[H]
    \resizebox{\textwidth}{!}{%
    \begin{tabular}{@{}p{2cm}l>{\raggedright}p{5cm}>{\raggedright}p{1.1 cm}>{\raggedright}p{1cm}>{\raggedright\arraybackslash}p{2cm}@{}} \toprule
        Fact-checking stages & No. & Design ideas & New idea & New need & Related AI research \\ \midrule
            % Forecasting claims & $1$ & Forecasting Disinformation & $\checkmark$ & $\checkmark$ & \cite{Huang2022-qh, Pan2023-me}\\ \midrule
            %\multirow{5}{2cm}{Monitoring claims} & $2$ & Identifying Broader Disinformation Narratives &  & $\checkmark$ & \cite{Jiang2021-ef, Hardalov2021-zm, Arakelyan2023-ju}\\ \cmidrule(r){2-6} 
            % & $3$ & Dynamic Credibility Monitoring of Social Media Users & $\checkmark$ & $\checkmark$ & \cite{zhou2020survey, Nguyen2018believeornot, Giachanou2019-qf, Zhi2017-wv, Rashkin2017-ct}\\ \cmidrule(r){2-6} 
            Monitoring claims & $12$ & Tracing the Spread of Similar Claims across Social Media Platforms &  &  & \cite{Kazemi2021-vg, Corney2021-sm, Shaar2020-eb}\\ \cmidrule(r){2-6} 
            & $13$ & Matching Similar Multilingual Claims &  &  & \cite{Koehn2017-bn, Wang2020-ie}\\ \cmidrule(r){2-6} 
            & $14$ & Detecting and Checking Claims from Multi-modal Content &  &  & \cite{Alam2021-zd, castro2022wild, wu2021towards, reddy2021detecting}\\ \cmidrule(r){1-6} 
            % \multirow{3}{2cm}{Selecting claims} & $4$ & Finding and Providing Context for Ambiguous Claims & $\checkmark$ &  & \cite{Choi2021-bm, Chen2022-yz}\\ \cmidrule(r){2-6} 
            % & $5$ & Personalized Claim Filtering and Selection &  & $\checkmark$ & \cite{alam2020fighting, Hassan2017TowardAF, vasileva2019takes, Nakov2022TheCC, alam2021fighting, Gupta2023-jx, Kartal2020-hp, Hardalov2021-zm, Arakelyan2023-ju}\\ \cmidrule(r){2-6} 
            % & $6$ & Personal Bias Warning System for Claim Selection & $\checkmark$ & $\checkmark$ & \cite{Hardalov2021-zm, Arakelyan2023-ju}\\ \cmidrule(r){1-6}
            % \multirow{3}{2cm}{Investigating facts} & $7$ & Human-AI Teaming for Fast Generation of Fact-check Briefs & $\checkmark$ & $\checkmark$ & \cite{Fan2020-sj, Chen2022-yz, Nguyen2020-we, Chen2023-fu}\\ \cmidrule(r){2-6} 
            % & $8$ & Identifying Official Databases and Formulating Queries for Verifying Quantitative Claims & & $\checkmark$ & \cite{Saeed2021-ff, Karagiannis2020ScrutinizerAM, Balalau2022-wu}\\ \cmidrule(r){2-6}
            Investigating facts & $15$ & Recommending Claim-Specific Experts &  &  & \cite{becerra2006searching, balog2012expertise}\\ \cmidrule(r){1-6}
            % \multirow{3}{2cm}{Writing fact-checks} & $9$ & Understanding Reader Engagement for Fact-checking Reports & $\checkmark$ & & \cite{Zhang2021-mh}\\ \cmidrule(r){2-6} 
            % & $10$ & AI Assistance in Writing Fact-checking Reports & $\checkmark$ & $\checkmark$ & \cite{Chen2023-fu}\\ \cmidrule(r){2-6} 
            % Writing fact-checks & $11$ & AI Assistance in Critiquing and Editing Fact-checking Reports  & & $\checkmark$ & \cite{Vo2019-mr}\\ \bottomrule
	\end{tabular}
	}
    \vspace{0cm}
    \caption{Four additional ideas generated that remain significant and merit further attention, despite earlier coverage in prior work.} \label{tab:all-ideas}
\end{table}

\subsection{\textbf{Tracing the Spread of Similar Claims across Social Media Platforms}} \label{idea12}

We learned from our participants that as online misinformation spreads, it becomes warped, eventually growing to encompass a plethora of similar, yet nonidentical, claims. Our participants want to group these similar, related claims together to gain a more accurate view of ``how these claims evolve [over time]'' (P1) and ``[to learn] the full scope of its virality on whatever platforms'' (P5). They (P12, 15, 20) pointed out two key functions of tracing similar claims: 1) retrieving checked claims similar to an unchecked claim at hand; and 2) grouping similar unchecked claims across different platforms.  For example, when deciding whether to pursue a new fact-check, our participants (P2, 9, 13-14, 18, 29) typically first investigate whether the claim has already been checked using rudimentary Google searches, to avoid overlap. However, participants may continue to pursue the claim despite the fact it is similar to the previously checked claims if certain criteria are met, such as: 1) it is highly relevant to their targeted readers (P2, 18); 2) it needs supplemental evidence to be fully fact-checked in a local context (P9, 13); or 3) its previous report does not meet high fact-checking standard (P18, 20)\footnote{The high standards of a fact-check includes its high readability and coherence. More details are described in the Appendix \ref{appendix:criteria-2}.}. 

This `provenance tracing' tool would enable fact-checkers to not only investigate whether a claim has been checked before, but also identify its provenance, tracking its spread across multiple platforms. Regarding specific features, P14 would like to sort claims by similarity relative to the ``primary'' claim. P9 and P12 wanted to order claims by the time of posting, so as to be helpful in investigating a claim’s provenance.

Claim matching could be directly applied to retrieve previously fact-checked claims that are lexically or semantically similar to a new claim \citep{Kazemi2021-vg, Corney2021-sm, Shaar2020-eb}. With the development of ClaimReview\footnote{\url{https://schema.org/ClaimReview}} - a tagging system for existing fact-checks - it is now much easier to retrieve previously checked claims to enable this function. As misinformation often backfires, researchers can also utilize natural language inference (NLI) to auto-verify these claims based on the veracity of similar previously fact-checked claims. Fact-checkers then can conduct a quick sanity check on these automated fact-checks and prioritize more time checking other claims. 

To aggregate similar unchecked claims, participants suggested that existing claim detection models could be combined with information retrieval \citep{Ryu2012-ef, Shao2016-af}, each fine-tuned on a specific media type for that platform. A claim could be searched for across all of these models and the results could be aggregated. Additionally, although there have been recent developments of APIs from different media platforms (e.g., Meta\footnote{\url{https://research.facebook.com/blog/2021/3/new-analytics-api-for-researchers-studying-facebook-page-data/}} and TikTok\footnote{\url{https://newsroom.tiktok.com/en-sg/an-update-on-our-platform-api-for-researchers-sg}}) and new cross-platform third-party retrieval tools (e.g., Tracking Exposed\footnote{\url{https://tracking.exposed/}}), participants requested more engineering work to combine these resources into a unified technological infrastructure for mass-monitoring checkable claims.

\subsection{\textbf{Matching Similar Multilingual Claims}} \label{idea13}

Misinformation is often spread through multiple languages, reaching larger populations through new languages. For our participants, it is important to catch these claims quickly, however, this becomes difficult without advanced knowledge of multiple languages. Our participants in multilingual countries, such as the Philippines (P16), South Africa (P4, 6-8, 13), Nepal (P5, 10-11), India (P21), and Mexico (P17), specifically indicated the need to detect similar claims across multiple languages.

P21 requested cross-lingual information retrieval \cite{nie2010cross} to: ``search [the claim] in Ukrainian and Russian at the same time, so that [they] get ...more results on the same topic.'' P10 and P11 also echoed similar needs for detecting checkable claims in Nepali. Five of the eight (62.5\%) participants who fact-checked multilingual claims mentioned that this should be an important feature for existing fact-checking tools. P17 reported that the claim detection tool he used (Facebook's tool) was developed to only find claims in English and is useful about 15-20\% of the time to find Spanish claims. This function could help our participants to ``trace the impact'' (P21) of a claim and gain a more holistic view of how the claim has spread (similar to Idea \ref{idea3}). Additionally, it would also ``save [us] a ton of time'' (P17) by cutting down manual human translation and conducting multiple searches in different languages. Participants (P5, 10-11) also reported that off-the-shelf translation tools lack adequate performance for low-resources languages (LRLs) \citep{Koehn2017-bn}. For example, P10 said, when translating English claims to Nepali using Google Translate, Nepali claims are interpreted very literally and lack the language-specific context she needs. These findings support results from the previous studies \citep{graves2018understanding, nakov2021automated}, suggesting the need for further engineering work in translating high-resources languages into LRLs. 

To implement this idea for high-resource languages, automatic translation models perform relatively well and can be used. For LRLs, one approach to applying transfer learning to train LRLs from a large monolingual language corpus (e.g., English) \citep{Wang2020-ie}. More generally, additional dataset or model development may be needed to enable effective support across languages. 

\subsection{\textbf{Detecting and Checking Claims from Multi-modal Content}} \label{idea14}

With an increased percentage of misinformation originating from media beyond text, participants want AI to help them better check claims from audio, photo, and video content. This idea has two primary aspects: 1) detecting checkable claims from multi-modal content; and 2) checking these claims by tracing their provenance\footnote {this idea refers to two fact-checking tasks: monitoring claims and investigating facts.}. We learned that identifying checkable claims from audio or video content is more challenging than from text. P18, who primarily checks claims on TikTok, mentioned video feeds on TikTok are curated based on user interests, and cannot be centrally monitored by fact-checkers. Additionally, due to the segmentation of existing monitoring tools (e.g., Facebook’s tool only monitors Meta-owned platforms, FullFact Alpha monitors news outlets, and TweetDeck monitors Tweets), our participants (P5, 9, 12, 14, 20-21) are forced to use multiple tools. They voiced that creating a single tool that sources claims across platforms would save time.

Because, as participants told us, claims extracted from videos or photos are often intentional misinterpretations, they usually begin their fact-checking process by tracing the provenance: finding the original photos or videos. For example, P3 said that ``a video [claim] is taken out of context, and you have to watch the whole video to understand [and to check it].'' To combat this, they imagined improving existing reverse video or image search with more advanced features, namely, retrieving the original video and mapping the claim to its original timestamp. Note that implementing such tools would require a combination of methods from computer vision, multimodal NLP, and information retrieval (see \citep{Cao2020-oi} for an overview). 

To identify this multi-modal misinformation, our participants proposed that AI could first transcribe it to text, then extract claims from the resulting transcripts using NLP. Some recent work \citep{Alam2021-zd} has approached claim detection by combining image data, such as a screenshot from a video or an individual picture, with text data. However, directly identifying claims from videos and audio remains under-explored. Researchers may need to construct new datasets that pairs textual claims with corresponding audio and video content, both short-form (TikTok and Instagram Reels) and long-form (YouTube). Recent work in video question answering \cite{castro2022wild}, long-form video understanding \cite{wu2021towards}, and detecting content patterns in podcasts \cite{reddy2021detecting} may be explored in order to develop new methods for this task. Additionally, mass-monitoring multi-modal content also requires infrastructure work, such as cross-referencing information from different social media platforms. 

\subsection{\textbf{Recommending Claim-Specific Experts}}\label{idea15}

A tool that can recommend ``experts'' for verification of findings related to a claim is helpful in the fact-checking process (P1-2, 5, 8-10, 15). Organizations tend to maintain a database of experts. Our participants mentioned that they either need to manually search through such a database or identify new experts. Sometimes it is difficult to fact-check a claim without access to an expert ``there are many claims that we find it very difficult to fact check [since we don't have access to experts]''(P10).
Implementing a focused information retrieval technique like \textit{expert retrieval}\cite{becerra2006searching, balog2012expertise} can be investigated in the context of fact-checking to address this issue.

\section{Fact-checker Needs and Co-Designed Ideas influenced by Regions and Organizational Interests} \label{appendix:criteria-3}
By mapping participant criteria (Appendix \ref{appendix:criteria-1} and \ref{appendix:criteria-2}) to their demographics and organizations, we uncovered potential regional patterns that appear to influence participant needs and the co-creation of ideas. Nonetheless, these findings are preliminary and necessitate further systematic validation by subsequent research. We report them below.

First, we found that participants from Philippines, South Africa, Nepal, and New Zealand, predominantly concentrated on verifying local misinformation, indicating a strong need for personalization shared across different ideas. For example, participants (P5, 9, 11) emphasized the significance of scrutinizing regionally targeted misinformation. Because this misinformation could inflict direct harm upon local communities even if its virality is limited. On the other hand, participants working in international fact-checking organizations (P20, 21), although geographically distributed, engaged in global fake news by collaborating with international colleagues where the claimants or subjects of fake news are located. This variance in regional focus, whether local or global, indicates a common interest in tailoring claim filtering and selection (Idea 5, Section \ref{idea5}) and in examining the demographic composition of the readership for published fact-checks (Idea 9, Section \ref{idea9}). 

Also, the difference in regional political discourse and cultural context tends to shape what and how fact-checkers check. For example, P3 reported that because politicians in the country have learned to modify their tactics to circumvent fact-checking scrutiny, sometimes they also check vague claims from politicians. This situation requires them to use more nuanced ratings to assess the truthfulness of such claims. Consequently, she expressed a strong interest in finding and providing context for ambiguous claims (Idea 4, Section \ref{idea4}). This idea is beneficial not only for highlighting inaccuracies in claims and pinpointing arguments requiring further investigation, but it also acts as a tool to aid novice fact-checkers in cultivating critical analytical skills, a vital competency for professional fact-checkers. 

Furthermore, regional difference highlights a language challenge in the fact-checking process. Participants from multilingual countries such as Mexico, Philippines, South Africa, Nepal, and India emphasized multilingual support. This need is particularly relevant in tracking the spread of claims (Idea 12, Section \ref{idea12}) and matching new claims with previously fact-checked claims (Idea 13, Section \ref{idea13}).

Beyond regional differences, the affiliations of different fact-checking organizations influence the selection criteria for fact-checkers. While a majority of participants focused on debunking false news, few also check potentially true stories. Reported by editor participants (P7, 19), checking fake news is financially beneficial for their organizations, particularly in collaboration with mainstream social media platforms like Facebook, helping them identify online hoaxes, fake news, or propaganda. Conversely, P18 (who is primarily engaged in video fact-checking for cable news) mentioned a preference for choosing stories based on viewer interest, regardless of their perceived veracity. These differences highlight the need for multifaceted features in tools designed for claim filtering and selection (Idea 5, Section \ref{idea5}). Additionally, there is a recognized need to refine existing annotation schema and techniques for claim detection to adapt to diverse contexts of claims-making. 

The editorial process for writing fact-checks also varies among organizations. Smaller entities with limited resources, like those represented by participants (P5, 10, and 11), rely on peer-review to edit their fact-check reports. In contrast, larger and more established organizations often implement a more rigorous process, involving two or three rounds of editing. Despite these differences, participants from both types of organizations were interested in using AI assistance to critique and edit fact-check reports (Idea 11, Section \ref{idea11}). This use of AI is seen as a way to compensate for limited editorial resources or to enhance the efficiency and productivity of the editing process.

%TC:endignore
\end{document}